\documentclass[aoas,preprint]{imsart}

\RequirePackage[OT1]{fontenc}
\RequirePackage{amsthm,amsmath}
\RequirePackage{natbib}
\RequirePackage[colorlinks,citecolor=blue,urlcolor=blue]{hyperref}
\usepackage{graphicx}
\usepackage{graphicx} 
\usepackage{hyperref} 
\usepackage{amsmath}
\usepackage{multirow}
\usepackage{subfigure}
\usepackage{latexsym, amssymb, amsbsy, epsfig, enumerate}

\def\calR{{\mathbb R}}  
\def\b0{{\textbf 0}}

\def\bt{{\textbf t}}

\def\bv{{\textbf v}}

\def\bx{{\textbf x}}

\def\bA{{\textbf A}}

\def\bC{{\textbf C}}
\def\bD{{\textbf D}}

\def\bG{{\textbf G}}
\def\bK{{\textbf K}}
\def\bI{{\textbf I}}
\def\bM{{\textbf M}}
\def\bS{{\textbf S}}
\def\bQ{{\textbf Q}}

\def\bX{{\textbf X}}
\def\bY{{\textbf Y}}

\def\bbeta{\boldsymbol{\beta}}

\def\bSigma{\boldsymbol{\Sigma}}

\def\btheta{\boldsymbol{\theta}}
\def\btheta{\boldsymbol{\theta}}

\startlocaldefs
\numberwithin{equation}{section}
\theoremstyle{plain}

\endlocaldefs

\begin{document}

\begin{frontmatter}
\title{Modelling ocean temperatures from bio-probes under preferential sampling}
\runtitle{Preferential Sampling}

\begin{aug}
\author{\fnms{Daniel} \snm{Dinsdale}\ead[label=e1]{danielrdinsdale@gmail.com}}
\and
\author{\fnms{Matias} \snm{Salibian-Barrera}\ead[label=e2]{matias@stat.ubc.ca}}


\affiliation{The University of British Columbia} 

\address{Department of Statistics\\
Faculty of Science\\
3182 Earth Sciences Building, 2207 Main Mall\\
Vancouver, BC Canada V6T 1Z4 \\
\printead{e1}\\
\phantom{E-mail:\ }\printead*{e2}}

\end{aug}

\begin{abstract}
In the last 25 years there has been an important increase in the amount of 
data collected from animal-mounted sensors (bio-probes), which are often used to study
the animals' behaviour or environment. We focus here on an example of the latter,
where the interest is in sea surface 
temperature (SST), and measurements are taken from sensors mounted 
on Elephant Seals in the Southern
Indian ocean. We show that standard geostatistical models may not be reliable
for this type of data, due to the possibility that the regions visited by the animals 
may depend on the SST. 
This phenomenon is know in the literature as preferential sampling, 
and,
if ignored, it may affect the resulting spatial predictions and parameter estimates. 
Research on this topic has been mostly restricted to stationary sampling 
locations such as monitoring sites. 
The main contribution of this manuscript is to extend this methodology to 
observations obtained by devices that
move through the region of interest, 
as is the case with the tagged seals. 
More specifically, we propose a flexible framework for inference on preferentially 
sampled fields, where the process that generates the sampling locations is stochastic and  
moving over time through a 2-dimensional space. Our simulation studies confirm that
predictions obtained from the preferential sampling model are more reliable
when this phenomenon is present, and they compare very well to
the standard ones when there is no preferential sampling.
Finally, we note that the conclusions of our analysis of the SST data can
change considerably when we incorporate preferential sampling in the model. 
\end{abstract}



\end{frontmatter}

\section{Introduction}\label{sec:intro}
The use of animal mounted sensors (bio-probes) to 
analyse population patterns has grown quickly
in the last 25 years~\citep{Fedak2004, Ungar2005, Evans2013}, with tags attached to both marine 
and land based 
animals. These tags can be used to provide valuable information by collecting data on the environment where the animals live, particularly in regions that are difficult to observe otherwise. 
One example is given by the use of marine mammal tags to measure oceanographic data, 
such as water temperature, salinity and others. \cite{Fedak2013} highlight the usefulness of such tags in 
profiling oceanographic data in polar regions, where data is typically difficult to obtain.

Although the methodology described in this paper is applicable to a range of different problems, we will focus here on data collected from CTD (Conductivity-Temperature-Depth) bio-probe tags attached to Elephant Seals in the Southern Indian ocean. These data were collected and made freely available as part of the MEOP (Marine Mammals Exploring the Oceans Pole to Pole) database~\citep{Roquet2013} and we utilise the South Indian ocean data subset described by \cite{Roquet2014}. This data set was collected to supplement the Advanced Research and Global Observation Satellite (Argos) float and ship based measures of water masses in typically under sampled areas of the Southern oceans, an area which is drastically changing and needs to be further understood~\citep{Jacobs2006}. While Northern oceans have been regularly sampled since the early 2000s using Argo profilers~\citep{Gould2004}, utilising this method in the Southern oceans is typically complicated by the presence of sea ice. 

\begin{figure}
\centering     
\includegraphics[width=4.3in]{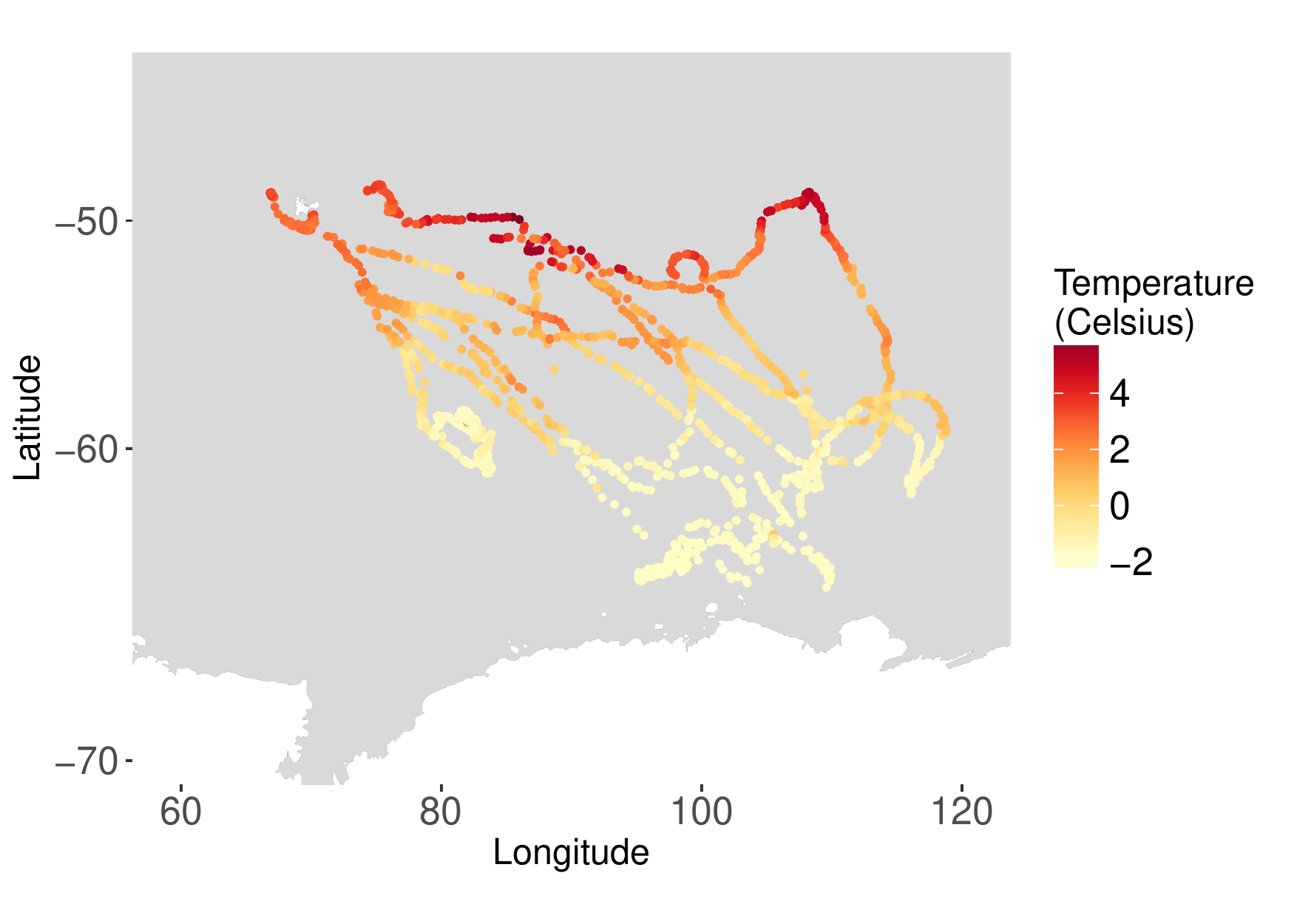}
\caption{Plot of all 1630 observations from the subset of Southern Elephant seal data analysed in Section~\ref{sec:realdata}. The first observation was taken on July 1st 2012 and the final observation on 30th September 2012.}
\label{fig:CoriolisEastFull}
\end{figure}

The data consist of location coordinates (longitude and latitude which are only available when the animal is surfaced) and corresponding sea surface temperature (SST) measurements. The animal locations in our applied example are determined using Argos and are typically accurate within $\pm5$ kilometers, whilst the temperatures are accurate within $\pm0.03^\circ$C~\citep{Roquet2014}. We use only the temperature data recorded at a depth of 6 meters, to represent the SSTs as closely as possible, and restrict ourselves to the region between -45 and -65 degrees latitude, between 60 and 120 degrees in longitude over the months of July to September 2012 and use tracks with 50 observations or more. The final data set consists of 9 separate tracks with 1630 observations in total, which can be seen in Figure~\ref{fig:CoriolisEastFull}.

Geostatistical models and methods~\citep{Diggle2007} provide a natural framework to analyse these data.  
While standard spatial statistical methods consider sampling locations that 
are chosen independently from the response variable of interest, it is important to note that 
the movement of the tagged animals 
(and thus, the locations at which measurements are taken)
may depend on 
the variable of interest (SST). Elephant Seals are likely to adjust 
their foraging due to the warming of ocean temperatures~\citep{Mcintyre2011}, 
because their prey tends to prefer
cooler and deeper waters. 
It also appears that there are less successful forages when diving in warmer water~\citep{Guinet2014}. 

Preferential sampling~\citep{Diggle2010} refers to the situation where
the process that determines the data locations and the 
spatial field of interest may depend on each other
(as can be the case with data collected from animal-mounted tags).
%
The effect of preferential sampling on subsequent inference can be two-fold. 
On the one hand, 
data collected in this way may tend to include a reduced 
range of the response variable. For example, if seals follow prey towards colder
water, their water temperature measurements will tend to not include observations from warmer
regions of their environment. On the other hand, the locations at which data are collected
might carry useful information about the variable of interest. In the previous example,  
one may expect that regions that were not sampled had 
slightly warmer waters than those visited by the seals.  
The first issue refers to the fact that the sample of responses may not be a reliable representation over the area of interest, 
while the second one implies that it may be advantageous for the analysis to 
take into account the observed locations rather than conditioning on them, as it is done
in standard spatial statistical methods. 

The impact of preferential sampling on estimation and prediction 
has been discussed in various recent papers. \cite{Shaddick2014} 
highlighted the preferential nature by which air pollution monitoring sites 
were added and removed from a UK monitoring network from the 1960s until 2006. 
Particularly in the early years of the study, there is evidence that removed sites had a 
lower annual mean pollution reading than those which were added.  \cite{Gelfand2012} 
showed the prediction effect of preferentially chosen ozone monitoring sites in California, 
whilst \cite{Diggle2010} and \cite{Dinsdale2017} illustrated the effect that preferential sampling of 
lead concentration in Galicia may have on the resulting predictions. \cite{Pati2011} studied the effect of preferential selection of monitoring sites measuring ozone levels in Eastern U.S.A. and more recently \cite{Conn2017} showed that preferential sampling in animal population surveys may cause large biases in the animal density estimates, giving an example using aerial survey counts of bearded seals in the Eastern Bering sea.

Research on the issue of preferential sampling has so far mostly been restricted 
to stationary sampling locations such as monitoring stations. 
The main contribution of this manuscript is to extend this methodology to 
observations obtained by devices that
move through the region of interest, such as those mounted on animals or people.
In this paper the animal locations are modelled 
using a correlated random walk \citep{Jonsen2005, Johnson2008}, 
where, 
to allow for the animal movement to depend on the 
variable of interest, we allow the drift function (which represents the direction preference at 
each location) to depend on the SST field. 
If this field can be assumed to be differentiable  
we include a term depending on its gradient to 
account for the animals' possible preference for particular water temperatures. 
Finally, we use a behavioural state component as in \cite{Auger2017}
to allow for a combination of ``momentum'' and environmental preferences 
in the animal's movement. 
Although we believe this movement model to be 
useful for the analysis of the CTD data, the focus of our work 
is on studying how to incorporate the possibility of
preferential sampling to this type of models, rather than advocating for 
the superiority of a specific movement model over others. 

It is interesting to note that the effect of preferential sampling
in these ``dynamic'' 
spatial models (where sampling locations move through the domain) 
can be different from what happens when sampling locations are 
stationary. In the latter case, when there is no
preferential sampling, observed locations are usually assumed to have a non-informative point pattern sampling distribution over the spatial domain. Thus, when preferential sampling is present it often results in distinct and potentially informative patterns in the 
monitoring locations \citep{Diggle2010, Dinsdale2017}. In contrast, even without 
preferential sampling, the locations visited by the animals
in our study would typically not be expected to be evenly
distributed over the area of interest (due to the 
dynamic nature of their movement). As a result, the 
information about SST contained in the locations the animals visited
(or in the regions without observations) may be less apparent for these
dynamical models than it is for stationary locations. Nevertheless, we show below that not
taking into account preferential sampling may still negatively affect the quality
and reliability of the estimated parameters and predictions. 

To estimate the parameters of interest in the model, we utilise a
likelihood approach based on a standard discretisation of the 
movement equations. The dependence between the 
sampling locations and the 
field of interest (SST) results in a likelihood that is
computationally challenging to optimize. 
We follow \cite{Dinsdale2017} in using the flexible Template Model 
Builder \texttt{R} package \texttt{TMB}~\citep{Kristensen2016} 
to deal with the computational complexity of 
the preferential sampling analysis. 

In many related models discussed in the literature, the underlying
field is only assumed to be continuous (but not differentiable). 
This choice appears to originate on mathematical and computational complexity
considerations rather than on the specifics of the phenomenon being studied
\citep{Rue2009, Lindgren2011, Simpson2012}. 
Since our model
involves the gradient of the underlying SST field
(see equation \eqref{eq:expecteddrift}), we will consider 
differentiable Gaussian random fields, which are 
relatively easy to use in the likelihood function when the latter
is approximated using using \texttt{TMB}. 
Details can be found in Appendix~\ref{appendix:smoother}. 


%
%
%


Our numerical experiments confirm that better predictions and parameter
estimates can be obtained when the model appropriately reflects 
the potential presence of preferential sampling. Moreover, if
the sampled locations are not preferentially chosen, the resulting 
predictions and parameter estimates are very close to those 
obtained with the standard model that conditions on the locations. 
Interestingly, our analysis also shows that the predicted SST in our
motivating example obtained with and without a preferential sampling model 
are in fact different, which suggests that the former may be more reliable. 

The rest of the paper is organized as follows. 
Section~\ref{sec:prefsampling} briefly reviews preferential 
sampling spatial models and the methods used to perform inference and prediction based on them.  
Our model for the SST data 
is discussed in Section~\ref{sec:SDEs}. 
The simulations results reported in Section~\ref{sec:simulateddata} 
illustrate the benefits of accounting for preferential sampling
when it may be present. 
%
Section~\ref{sec:realdata} contains the analysis of Southern Indian ocean temperatures from tagged Elephant Seals, where we compare how 
accounting for possible preferential sampling might alter the conclusions reached by researchers.
A final discussion can be found in Section~\ref{sec:disc}.

\section{The Preferential Sampling Problem}\label{sec:prefsampling}

\subsection{Standard Model Framework}

Our Elephant Seal data consists of a response variable of interest (SST with possible measurement error, which we will denote by $Y$), a time stamp and the 
corresponding location in latitude and longitude coordinates $(\bX \in \calR^2)$ which we assume contains no measurement error, see Section~\ref{sec:assumptions} for a discussion on this assumption. 
Since these tags measure water temperature at regular intervals, 
we consider the temperature measurement taken immediately before 
a location
was obtained (which happens when the animal surfaces). 
We assume that the measurements taken from each trip
are independent from each other.


%
To model these data we follow the geostatistical framework 
and notation 
of~\cite{Diggle2007}.
More specifically, we assume that the data consist
of a finite set of observations from a spatially continuous 
phenomenon $\left\{S(\bx):\bx\in\mathcal{D}\subseteq\mathbb{R}^2\right\}$.
In our application, $S(\bx)$ denotes the true SST at location $\bx$, and ${\cal D}$ is the region of the 
Southern Indian ocean. Note that $S(\cdot)$ does not vary over time, we discuss this further in Section~\ref{sec:assumptions}. 
The model for the measurements $Y_1$, \ldots, $Y_n$ obtained in one trip is:
\begin{equation}\label{eq:geoframework}
Y_i \, = \, \mu+ S(\bX_i) + Z_i, \quad \bX_i \in \mathcal{D} \, , \quad i=1, \ldots, n \, ,
\end{equation} 
where $\mu\in\mathbb{R}$ is a constant mean parameter over $\mathcal{D}$ and $\bX_i$ is the measurement location. The $Z_i$ is included in the model 
above to account for measurement errors in the sea surface temperatures, and are assumed to be
mutually independent random variables with mean 0 and so called ``nugget variance'' $\tau^2$. 

We assume that the SST field $S$ is a Gaussian Process with mean 0 and Mat\'{e}rn covariance function
given by 
\begin{equation}\label{eq:matern0}
C(r) =  \sigma^2 \, \frac{2^{1-\kappa}}{\Gamma(\kappa)}\left(\frac{r}{\phi}\right)^{\kappa}K_{\kappa}\left(\frac{r}{\phi}\right), \quad r>0,
\end{equation}
where $r$ is the distance between two points and $K_{\kappa}$ is the modified Bessel function of the second kind. 
The scale (range) $\phi$ and smoothness $\kappa$ parameters control the rate of correlation 
decay over distance and $\sigma^2$ the marginal variance of the process (partial sill). Furthermore, $\kappa$ controls the smoothness of the process realizations. 
Since our model 
involves the gradient $S$ (see \eqref{eq:expecteddrift}), in what follows we will 
assume that the surface of SST is differentiable (in mean-square sense), 
which corresponds to $\kappa> 1$~\citep{Diggle2007}. More specifically, 
we take $\kappa = 2$ which results in a spatial process $S$ that is mean-square 
differentiable~\citep{Banerjee2003, Banerjee2003b}. 

The main goal of our analysis is to obtain predictions for the SST 
field $S$ over a grid of unobserved locations. Given estimates for the unknown
field parameters in the model 
one can use 
standard spatial prediction techniques (e.g. kriging). 
However, as discussed in \cite{Diggle2010} and \cite{Dinsdale2017}, when the 
locations $\bX$ may depend on the field of interest 
$S$, one can obtain better predictions by 
including the information on $S$ contained in $\bX$.
In the rest of this section we show how to construct an appropriate likelihood
function that incorporates the possibility of preferential sampling,
and how it can be maximized to obtain more accurate parameter estimates and 
predictions for the field $S$. 

%

\subsection{Maximum Likelihood Estimation}\label{sec:likelihood}
We use $[\bA; \btheta]$ to denote the density or probability mass function of the random object $\bA$, which depends on a vector of parameters $\btheta$. 
In our case, $\btheta$ is the vector of all parameters in the model, which can be partitioned as
$\btheta = (\btheta_{F}^\top, \btheta_{L}^\top)^\top$, where $\btheta_{F}$ are the parameters of the latent field process (e.g. $\btheta_{F}=(\mu, \tau, \kappa, \phi, \sigma^2)^\top$ when using the model from \eqref{eq:geoframework} and \eqref{eq:matern0}), and  $\btheta_{L}$ are the parameters of the sampling location distribution. This last vector is divided into $\btheta_{L1}$, which are parameters that account for dependence between sampling locations and the latent field, and $\btheta_{L2}$, which do not relate to this dependence: $\btheta_{L} =( \btheta_{L1}^\top, \btheta_{L2}^\top )^\top$. 

We then consider the likelihood function $L(\boldsymbol{\theta})$ 
based on the observed data $\bY$ and $\bX$:
\begin{equation} \label{eq:likelihoodorig}
L(\boldsymbol{\theta}) \, = \, \left[ \bX, \bY ; \btheta \right] \, = \, 
\int \left[\bX, \bY, S ; \btheta \right] \mathrm{d}S \, .
\end{equation}
Typically, one has $\left[ \bX, \bY, S; \btheta \right] =
\left[ \left. \bY \right| S, \bX; \btheta_F \right] \left[ \left. \bX \right| S ;\btheta_L \right] \left[ S; \btheta_F \right]$. 
Standard geostastical models assume that the process that selects the 
measurement locations $\bX$ is independent
from the response process $S$ (in symbols: $\left[ \left. \bX \right| S; \btheta_L \right] = \left[ \bX; \btheta_{L2} \right]$), and
hence $\left[ \bX, \bY, S; \btheta \right]  =
\left[ \left. \bY \right| S, \bX; \btheta_F \right] \, \left[  \bX; \btheta_{L2} \right] \, \left[ S; \btheta_F \right]$. 
In this case it follows that $L(\boldsymbol{\theta}) = \left[ \bX; \btheta_{L2} \right] \left[ \left. \bY \right| \bX; \btheta_F \right]$,
and inference about $\btheta_{F}$ can be carried out conditionally on the observed locations $\bX$. 


Preferential sampling refers to the situation where the observed locations $\bX$ may depend 
on the unobserved process $S$. 
When $\left[ \left. \bX \right|  S; \btheta_L \right] \neq \left[ \bX; \btheta_{L2}\right]$ 
care must be taken when constructing the likelihood function 
in \eqref{eq:likelihoodorig}. In this case we cannot simply condition 
on the sampling locations, but should rather use the full likelihood function:
\begin{equation}\label{eq:jointdistint}
L(\boldsymbol{\theta}) = 
\int \left[ {\bX}, \bY, S; \btheta\right]\mathrm{d}S \, = \, 
\int \left[ \left. \bY \right| S, \bX; \btheta_F \right]  \left[ \left. \bX  \right| S; \btheta_{L}\right]  \left[ S; \btheta_{F}\right] \mathrm{d}S \, .
\end{equation}

%
%


\subsection{Preferential Sampling Using Template Model Builder} \label{sec:prefTMB}
Evaluation of the integral \eqref{eq:jointdistint} is computationally challenging and 
hence optimising the function is a difficult problem. 
~\cite{Diggle2010} proposed a Monte Carlo (MC) approximation to a discrete 
version of the integral, namely
\begin{equation}\label{eq:jointdistintdiscrete}
\int \left[ \left. \bY \right| \bS, \bX ; \btheta_F \right]  \left[ \left. \bX  \right| \bS; \btheta_L \right]  \left[ \bS; \btheta_F \right] \mathrm{d}\bS,
\end{equation}
where $\bS$ is a set of values of $S$. The exact locations used in the discretisation depends on the model used for the sampling locations and is discussed in more detail in Section~\ref{sec:moveint}. 
A direct MC approximation via simulated instances of the vector $\bS$ is particularly inefficient, since many
of the realisations of $\bS$ may not be compatible with the observed 
measurements $\bY$. 
Alternative representations of the likelihood function
require sampling from the distribution of the discretized field $\bS$ conditional on the observed locations
and measurements (i.e.\! $\bS | \bY, \bX$), which is generally intractable~\citep{Dinsdale2017}. 
%
%
%
%
Other alternatives to inference under preferential sampling have been proposed. For
example, \cite{Pati2011} considered a Bayesian alternative and one could also use the \texttt{R-INLA} 
package~\citep{Rue2009}, which utilises integrated nested Laplace approximation. 

A common assumption made in the literature about preferential sampling is that,
conditional on the random field $S$, the sampling locations are static, often modelled via an 
inhomogeneous Poisson or similar process where the intensity function of $[\bX|S; \btheta_L]$ 
depends on $S$. 
However, in our application the sampling locations are obtained from a process 
continuously moving through the 2-dimensional domain. 
For this reason, we wish to use a flexible modelling framework in which we can evaluate the likelihood 
\eqref{eq:jointdistint} efficiently 
for more complex forms of $[\bX|S; \btheta_L]$. Although \texttt{R-INLA}  provides a highly efficient computational framework, it did not accommodate our relatively complex models for $[\bX|S; \btheta_L]$
with an underlying smooth process $S$ (e.g. a mean square differentiable SST surface). 

We use 
the \texttt{R} package Template Model Builder (\texttt{TMB})~\citep{Kristensen2016} 
to maximise \eqref{eq:jointdistint} for the dynamic movements models discussed in the next Section.  
This package uses Automatic Differentiation (AD)~\citep{Griewank2008} of a Laplace Approximation to 
the likelihood to efficiently maximise it with respect to the full parameter vector $\btheta$. 
We define the joint negative log-likelihood function
\begin{equation}\label{eq:TMBlike}
f(\bS, \boldsymbol{\theta}) = -\log\left(\left[ \left. \bY \right| \bS, \bX ; \btheta_F \right]  \left[ \left. \bX  \right| \bS; \btheta_L \right]  \left[ \bS; \btheta_F \right]\right),
\end{equation}
and \texttt{TMB} computes an approximation to $ \int \exp[ -f(\bS, \btheta) ] \, \mathrm{d}\bS$, which can 
be optimized numerically with respect to $\btheta$. 


The dimension of the integral in \eqref{eq:jointdistintdiscrete} grows rapidly with the size of the
grid that is used to discretize the field $S$. Important efficiencies can be obtained by using 
the stochastic partial differential equation (SPDE) approximations for Gaussian fields~\citep{Lindgren2011},
which are also exploited by \texttt{R-INLA}. More specifically, these SPDE approximations 
allow the use of sparse precision matrices to more
efficiently evaluate the high dimensional integral in \eqref{eq:jointdistintdiscrete}.
Although the \texttt{R-INLA} package currently only allows 
continuous but not
differentiable fields $S$ in \eqref{eq:jointdistint}, it is not difficult to extend the same approach
for smoother random fields when using \texttt{TMB} to approximate \eqref{eq:jointdistintdiscrete}. 
In particular, we work with mean-square differentiable random 
fields~\citep{Banerjee2003, Banerjee2003b}. 
Details can be found in Appendix~\ref{appendix:smoother}. 

An important goal of this type of analyses is the prediction of SST on nearby 
locations that were not sampled. To construct predictions that take into account the 
preferential nature of the data, it is generally not sufficient to 
use kriging, even with parameter estimates obtained through a corrected likelihood 
function as in \eqref{eq:jointdistint}. Such an approach would effectively ignore the 
dependency between $\bX$ and $\bS$~\citep{daSilva2015, Dinsdale2017}. Although the true predictive distribution of $\bS$ is 
intractable in most cases when $\bX$ and $\bS$ are dependent, 
\texttt{TMB} provides point predictions and prediction variances from the
estimated mode of $[ \bS | \bY, \bX; \btheta]$ at 
$\btheta=\btheta_{\text{opt}}$, where $\btheta_{\text{opt}}$ is the 
vector of parameters that 
maximises \eqref{eq:jointdistintdiscrete}. Specifically, let
\begin{equation}\label{eq:TMBpointpred}
\hat{\bS}(\btheta) = \underset{\bS}{\operatorname{arg \, min}} f(\bS, \btheta) \, ,
\end{equation}
then $\hat{\bS}(\btheta_{\text{opt}})$ is a predictor for the 
discretised version $\bS$ of the process $S$ 
based on the preferential sampling model \eqref{eq:jointdistint}. 

\subsection{Assumptions}\label{sec:assumptions}
There are two key assumptions made so far in this section that should be noted. First is the assumption that the observed sampling locations $\bX$ are the true positions of the animals. In reality there will be some degree of measurement error attached to the sampling locations, which depends on the tag type. Argos locations tend to include significant measurement error in comparison to Global Positioning System (GPS) locations which is common in land based tracking such as polar bears~\citep{Auger2016} and various birds such as albatrosses~\citep{Weimerskirch2002} and gannets~\citep{Votier2010}. However such systems are less appropriate for marine systems since GPS requires several seconds of exposure to obtain a location estimate~\citep{Dujon2014}. Recently Fastloc-GPS tags (\url{http://www.wildtracker.com}) have become more popular due to their improved accuracy compared to Argos and that these systems only require a fraction of a second to obtain a location estimate. When using such data \cite{Auger2017} consider the location measurement error negligible enough to be ignored.

For the purpose of this paper, which is to emphasize the preferential sampling problem, we also decided to ignore measurement error in order to provide realistic but not overly complex models. However, this particular compromise between model complexity and computational efficiency by neglecting measurement error may have a serious impact on the analysis in certain situations. In particular, large unaccounted sampling location errors may lead to erroneous conclusions on the animal movement and consequently the preferential sampling effect. A possible strategy to incorporate measurement errors in the locations into our model is to include a latent state of true but unobservable locations (see for example \cite{Albertsen2015, Johnson2008}), which would add further latent states to the integral in \eqref{eq:likelihoodorig}. Note that in this case, care will be needed when considering the interplay between $S$ and $\bX$, since both objects will be unobservable.

The second assumption is that although the samples are taken at various time points, that SST depends only on location and not time. Therefore we can view the continuous SST field $S$ as constant over time. Further research in this area relaxing this assumption would be valuable. Enabling $S$ to vary over time, as it would in real life, would allow for analysis of data over longer periods of time with the model adapting to SST over various seasons and years. In this case one would consider the data to be of the form $Y_t=S(\bX_t, t)+Z_t$ so that $S(\cdot)$ is a function of both location and time. For the real data analysis in Section~\ref{sec:realdata} we consider data across only 3 months to reduce the impact of a changing temperature field.
\section{A Preferential Movement Model} \label{sec:SDEs}
To account for preferential sampling of ocean temperatures, we need to define a model for the location of marine mammals that takes into account possible relationships between movement velocity and ocean temperature. 
We will use discretised models as functions of the observed locations $\bX=(\bX(t_1),\ldots,\bX(t_n))$, in which movement may also depend on previous locations.
 In these cases, we can write the location at time $t_{k+1}$ as
\begin{equation}\label{eq:nonprefbasic}
\bX(t_{k+1}) = g_{\text{NP}}(\bX(t_{1:k}), \btheta_L) + \epsilon(\bt_{1:k}, \btheta_L)
\end{equation}
where $\bX(t_{1:k})=(\bX(t_{1}),\ldots,\bX(t_{k}))$ and $\btheta_L$ is now the vector of all movement parameters. The function $g_{\text{NP}}(\cdot)$ is some deterministic movement function where $\text{NP}$ stands for non-preferential and $\epsilon(\cdot)$ is an error term. 

Using the representation in \eqref{eq:nonprefbasic}, under preferential sampling we need to define a movement model of the form
\begin{equation}\label{eq:prefbasic}
\bX(t_{k+1}) = g_{\text{P}}(\bX(t_{1:k}), S, \btheta_L) + \epsilon(\bt_{1:k}, \btheta_L)
\end{equation}
where the function $g_{\text{P}}$ is now a function of $S$, therefore enabling the movement model to depend on the temperature field.


\subsection{A ``Preferential'' CRW Model for Marine Mammal Movement}\label{sec:CTCRWmarinemammals}
We consider a model similar to the first-difference correlated random walk (DCRW) model~\citep{Jonsen2005}. We wish to include non-regular time intervals to account for possible irregularity in the surfacing of the marine mammals. Maintaining a constant time step through interpolation of the data, as discussed in \cite{Jonsen2005, McClintock2012, Hooten2017Book} among others, is not possible in our preferential sampling framework. This is due to the necessity of maintaining the link between the sampling locations ($\bX$) and the corresponding latent field measurements ($\bY$). If we interpolate the trajectory it is not clear how we would obtain the corresponding $Y_i$ measurements at these interpolated locations, other than using a method such as kriging, which may dilute any preferential effect that was present in the original data. Another option may be to use thinning~\citep{Gurarie2017}, however we wanted to avoid this in our application in this paper, due to the limited temporal resolution of the data to which we have access. 

We term this model the ``preferential correlated random walk'' (PCRW) model and assume that the sampling locations $\bX(t_1),\ldots,\bX(t_n)$ follow 
\begin{equation}\label{eq:hybridSDEdiscrete}
\bX(t_{k+1}) = \bX(t_{k}) + \boldsymbol{\mu}(\bX(t_{1:k}), S, \btheta_L)(t_{k+1}-t_{k}) + \boldsymbol{\Sigma}(\btheta_L)\boldsymbol{A}_{k}\sqrt{t_{k+1}-t_{k}} \, ,
\end{equation}
where $\bA_k$ denotes a standard bivariate normal random vector, $\bSigma$ is a $2 \times 2$ matrix that corresponds to the variance of the diffusion terms, and $t_k$ are the observation times. 

To capture various movement patterns such as foraging and directed movement, rather than using discrete behavioural states~\citep{Morales2004, Breed2009, McClintock2012}, we propose a continuous behavioural state system similar to \cite{Auger2017, Breed2012}. This method was chosen to obtain a differentiable likelihood function through the Laplace approximation outlined in Section~\ref{sec:prefTMB}, which would be invalidated with the more commonly used discrete states~\citep{Bolker2013}. An alternative approach could be to estimate the movement parameters in \texttt{TMB}, then follow this with behavioural state estimation using the Viterbi algorithm~\citep{Whoriskey2017}.

The drift function and behavioural states at the measured locations and times satisfy:
\begin{equation*}
\boldsymbol{\mu}(\bX(t_{1:k}), S, \btheta_L) = f(\beta_{t_k})\boldsymbol{\phi}(\bX(t_{k}), S, \btheta_L) + (1-f(\beta_{t_k}))\bv(\bX(t_{1:k})) \, , \\
\end{equation*}
where $\beta_{t_k} \in \mathbb{R}$ for all $t_k$, $\bv:\mathbb{R}^2\to\mathbb{R}^2$ represents the ``velocity'' of the animal, and $\boldsymbol{\phi}:\mathbb{R}^2 \times \mathbb{R}^2 \to\mathbb{R}^2$ can be thought of as the foraging movement function that depends on the location and the latent temperature field $S$, which models the possible preference of the animals for different water temperatures. The behavioural state function  $f:\mathbb{R}\to [0,1]$ depends on $\beta_{t_k}$ and controls the auto-correlation of the movement at each time point. This ensures that when $f(\beta_{t_k})\approx 1$ then $\boldsymbol{\phi}(\bX(t_k) , S, \btheta_L)$ becomes the expected drift direction, whilst when $f(\beta_{t_k})\approx 0$ movement tends in the direction of the current velocity $\bv(\bX(t_{1:k}))$. 

It is important to note that the velocity function may depend on more than just the previous sampling location. Consequently, like the DCRW, our PCRW model is not Markovian. Although one may consider including a latent velocity state ($\bv$) similar to the continuous time correlated random walk model (CTCRW) of \cite{Johnson2008}, if this velocity was to depend on the locations $\bX$ and also $S$, such a model becomes drastically more complicated. We discuss this in more detail in Appendix~\ref{appendix:CTCRWdifficulties}. Alternative continuous-time models may also be adaptable to model preferential movement, however. For example, the correlated velocity model (CVM)~\citep{Gurarie2011} and functional movement models (FMMs)~\citep{Buderman2016, Hooten2017}. 

We consider a non-latent velocity state approximation, taken to be 
\begin{equation}\label{eq:velocityapprox}
\bv(\bX(t_{1:k})) = \frac{\bX(t_{k})-\bX(t_{k-1})}{ t_{k}-t_{k-1} },
\end{equation}
and specify a behavioural function
\begin{equation}\label{eq:behavfunction}
f(\beta_{t_k}) = \frac{\exp(\beta_{t_k})}{1+\exp(\beta_{t_k})}, \\
\end{equation}
so that as $\beta_{t_k}$ increases, so does the influence of $\boldsymbol{\phi}$, whereas when $\beta_{t_k}$ decreases the current velocity $\bv(\bX(t_k))$ becomes more of a factor in the movement. Therefore, our PCRW model can be written
\begin{align}\label{eq:PCRWdrift}
\begin{split}
\boldsymbol{\mu}(\bX(t_{k}), S, \btheta_L) &= \frac{\exp(\beta_{t_k})}{1+\exp(\beta_{t_k})}\boldsymbol{\phi}(\bX(t_{k}), S, \btheta_L) + \frac{1}{1+\exp(\beta_{t_k})}\bv(\bX(t_{1:k})) \, , \\
\beta_{t_{k+1}} &= \beta_{t_{k}} + \sigma_{\beta} B_{k}\sqrt{t_{k+1}-t_{k}},
\end{split}
\end{align}
where $B_k$ are univariate standard normal random variables and $\sigma_{\beta} > 0$ determines the evolution of the random states $\beta_{t_k}$. Note that the inclusion of the random $\beta$ states means we need to re-specify the preferential likelihood for our Laplace approximation, which we show in Appendix~\ref{appendix:newlik}. 
Initial values for $\bX(t_0)$ and $\beta_{t_0}$ used 
in our simulation studies are described in 
Appendix~\ref{appendix:datageneration}.


Depending on the application of the PCRW model, a variety of forms for the foraging function $\boldsymbol{\phi}$ may be used. In our case, to model the possible tendency of an animal to move towards particular water temperatures when searching for prey we propose the following:
\begin{equation}\label{eq:expecteddrift}
\boldsymbol{\phi}(\bX, S, \btheta_L) \, = \, -\alpha \, S(\bX) \, \nabla S(\bX) \, ,
\end{equation}
where $\alpha\in\mathbb{R}$, $S(\bX)$ is the value of the random field at location $\bX$ and 
$\nabla S(\bX)$ is the gradient of $S$ at $\bX$. Although the parameter $\alpha$ above may appear 
to be unidentifiable, this is in fact not the case when you consider the full likelihood 
function which also includes the density functions $[\bY|S,\bX; \btheta_F]$ and $[S; \btheta_F]$.

The form of $\boldsymbol{\phi}$ in \eqref{eq:expecteddrift} defines the expected drift as 
descending (or ascending if $\alpha<0$) along the gradient of the SST field, with a velocity that depends both on the 
temperature at the present location and a scalar $\alpha$. This is somewhat similar to the varying 
motility surface used by \cite{Russell2016} in an stochastic 
differential equation (SDE) model to allow the magnitude of the velocity 
vector to depend on the location. More specifically, we can view the latent field $S$ 
as a scaled potential surface in which the gradient of $S$ directs the expected 
movement with a velocity that also depends on the value of $S$ at that location. 
Potential surfaces have previously been used to model the movement of animals 
including monk seals~\citep{Brillinger2008}, elk~\citep{Brillinger2002,Preisler2013}, and ants~\citep{Russell2016}, with various estimation methods for potential surface SDE models compared by \cite{Gloaguen2018}.

It should be noted that appropriate forms of $\boldsymbol{\phi}$ for particular species may require specialist input and we do not claim that the one shown in \eqref{eq:expecteddrift} is necessarily the best model for seals. The preferential sampling effect may vary over species, locations and possibly even individuals. However, the form in \eqref{eq:expecteddrift} may identify preferential movement and we use it as an example for integrating a foraging function into the preferential sampling framework. Furthermore, in application we may wish to adjust \eqref{eq:expecteddrift} to 
\begin{equation}\label{eq:expecteddriftV2}
\boldsymbol{\phi}(\bX, S, \btheta_L) \, = \, -\alpha \, S^{*}(\bX) \, \nabla S(\bX) \, ,
\end{equation}
where $S^{*}=S+c$ for some constant $c\in\mathbb{R}$ specified by the user. We would do this to ensure that $S^{*}$ is the same sign across the domain, to prevent the switching of movement patterns when $S$ goes from negative to positive or vice versa.
%



Finally, using \eqref{eq:geoframework}, 
\eqref{eq:hybridSDEdiscrete}-\eqref{eq:expecteddrift} and a 
finite-differences approximation to the gradient of the field $S$
we construct a likelihood function that can be maximized
numerically, as described in Section \ref{sec:prefTMB} above.
Predicted values for the SST field 
can be obtained 
by using \eqref{eq:TMBpointpred} at the vector 
of optimal model parameters. 
%

\subsection{Relationship Between the Movement Model and Likelihood Integral}\label{sec:moveint}
In this section we discuss the form of the discrete grid $\bS$ used to approximate the integral in \eqref{eq:jointdistintdiscrete}. The locations at which we need to integrate over depends entirely on the distribution of $\bX|S$. Previous literature, in which the sampling locations were point patterns, suited a finely spaced lattice covering the entire domain~\citep{Dinsdale2017}. This was because every point in the domain was a factor in determining the distribution of sampling locations.

In the case of a moving animal measuring a temperature field, this may or may not be required. Take for example, a model that assumes the animal might be knowingly moving towards distant points of attraction that are related to the latent field, for example a high prey region with low water temperature. In this case, locations far away from the animal might impact on the movement, hence requiring an approach similar to the point pattern integrals in which we require using a finely spaced lattice covering the entire, or majority, of the domain. 

On the other hand, the preferential CRW model we are proposing in this paper assumes movement only depends on the animal's immediate vicinity. This can be seen by observing that the only influence of $S$ on the movement is in \eqref{eq:expecteddriftV2}, in which the current temperature and gradient of temperature field impacts movement. Hence, it would be more efficient to use a smaller grid for $\bS$, which contains only the sampling locations and those areas nearby which can be used to calculate the temperature gradient at sampled locations using a finite differences approach.

\section{Simulation Experiments}\label{sec:simulateddata}
In this section we discuss the results of a simulation study conducted 
to illustrate the effect of preferential sampling on the analysis of 
spatial data where sampling locations are moving through space, 
such as is the case in our SST data. 
The goal is to show to what extent incorporating preferential sampling 
in the model may improve the resulting SST predictions and parameter
estimates. 
Furthermore, our results indicate that when preferential sampling is not present, 
there is almost no difference between using a model that incorporates preferential 
sampling and the usual geostatistical model that conditions on the locations. The non-preferential analysis can be found in Appendix~\ref{appendix:nonpref}.

\begin{figure}[tp]
\centering     
\includegraphics[width=3.5in]{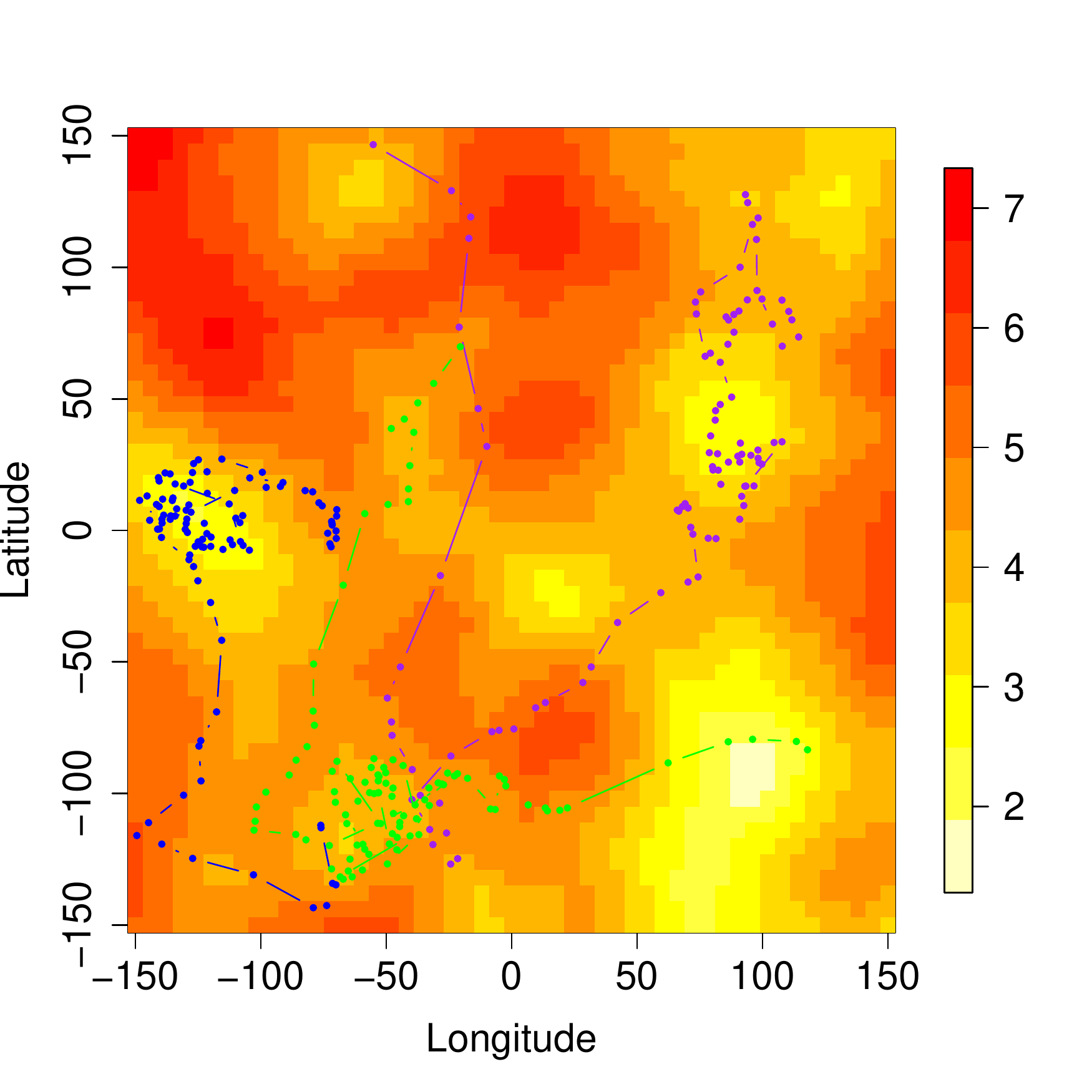}
\caption{Example of a simulated data set of 3 tracks 
generated using the field and movement parameters defined in Section~\ref{sec:simulateddata}, 
resulting in a moderate preferential sampling effect.}
\label{fig:hybridexample}
\end{figure}

We generated 100 data sets, each of them consisting of up to 300 observations
on 3 animal tracks following the Preferential-CRW movement model described in 
Section \ref{sec:CTCRWmarinemammals}. Data were first generated on a fine 
time grid, and a subsample selected to form each of the 100 data sets. 
Details on the data generation process be found in Appendix~\ref{appendix:datageneration} and the corresponding code found in \ref{suppA}.
With our simulation parameters we expect the tracks to oversample cooler regions. 
Figure~\ref{fig:hybridexample} shows one simulated data set 
where the preferential sampling effect is apparent. 


\subsection{Parameter Estimates}\label{sec:simest}
Note that because of the way the data was generated (by subsampling 
trajectories created on a relatively fine time scale), 
the estimated movement parameters may not correspond to their 
nominal values used to create the data~\citep{Gurarie2017}. Hence, 
we report here results for the estimates of the parameters of the 
spatial process.

\begin{figure}
\centering     
\subfigure[Mean ($\mu$)]{\includegraphics[width=2.3in]{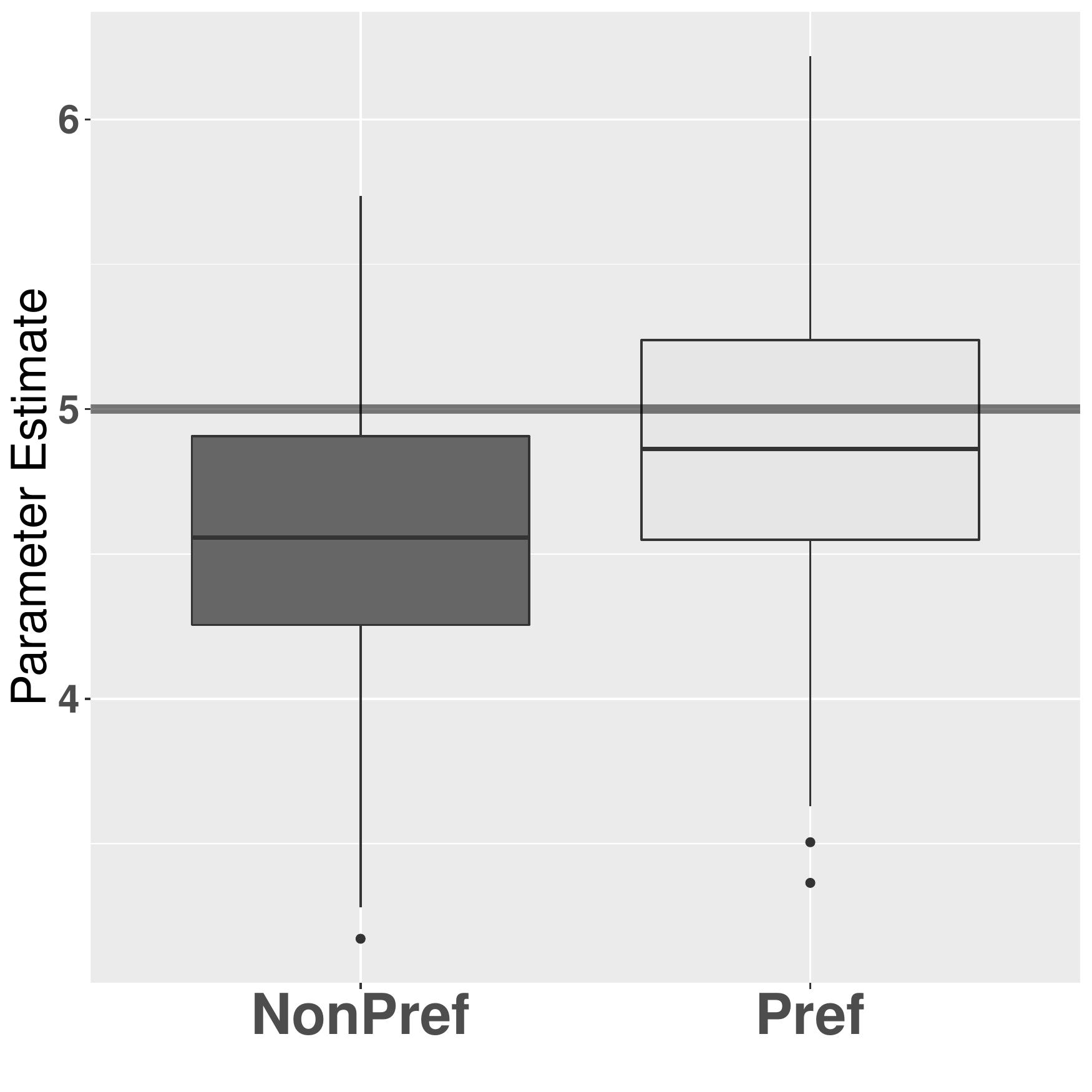}}
\subfigure[Scale ($\phi$)]{\includegraphics[width=2.3in]{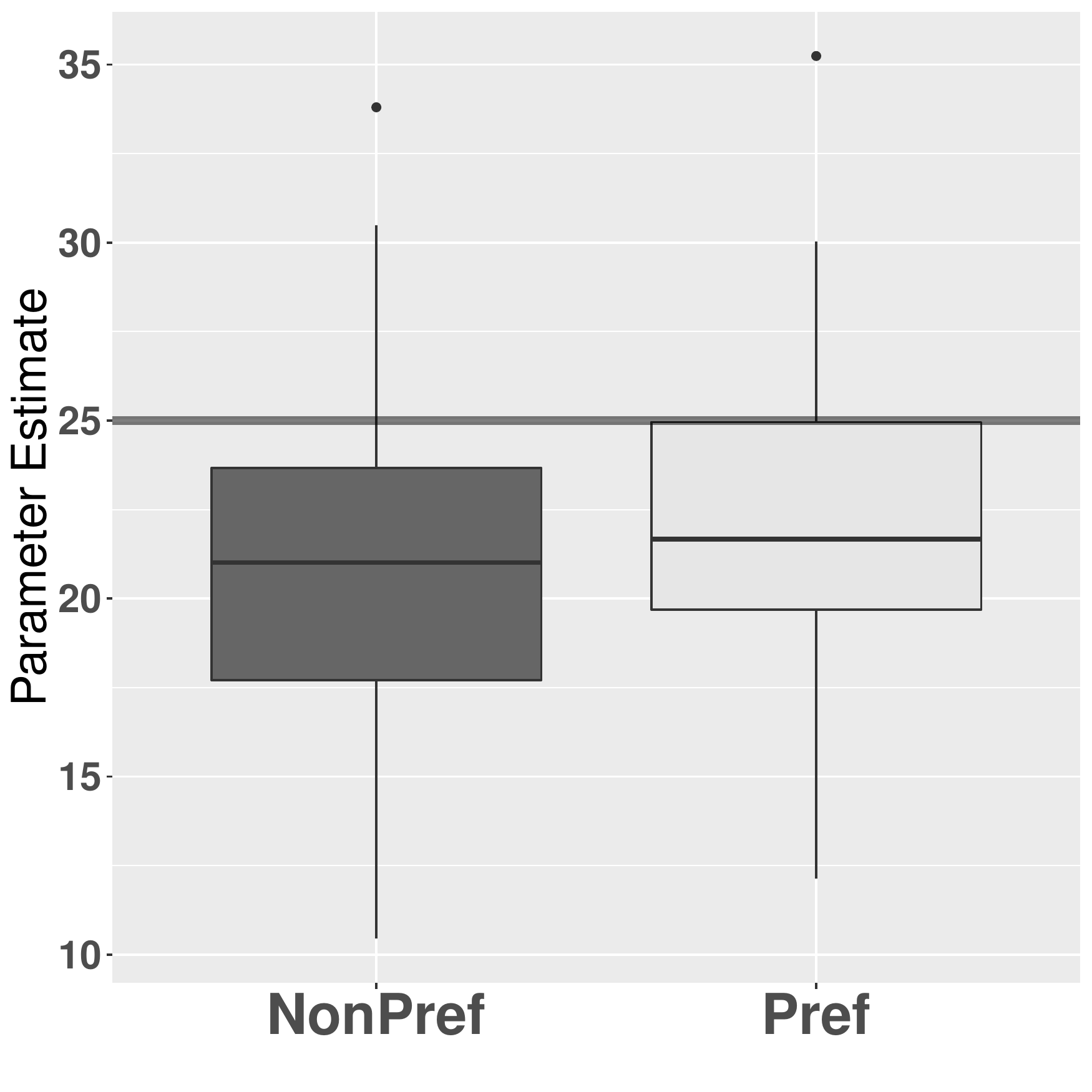}}
\subfigure[Variance ($\sigma^2$)]{\includegraphics[width=2.3in]{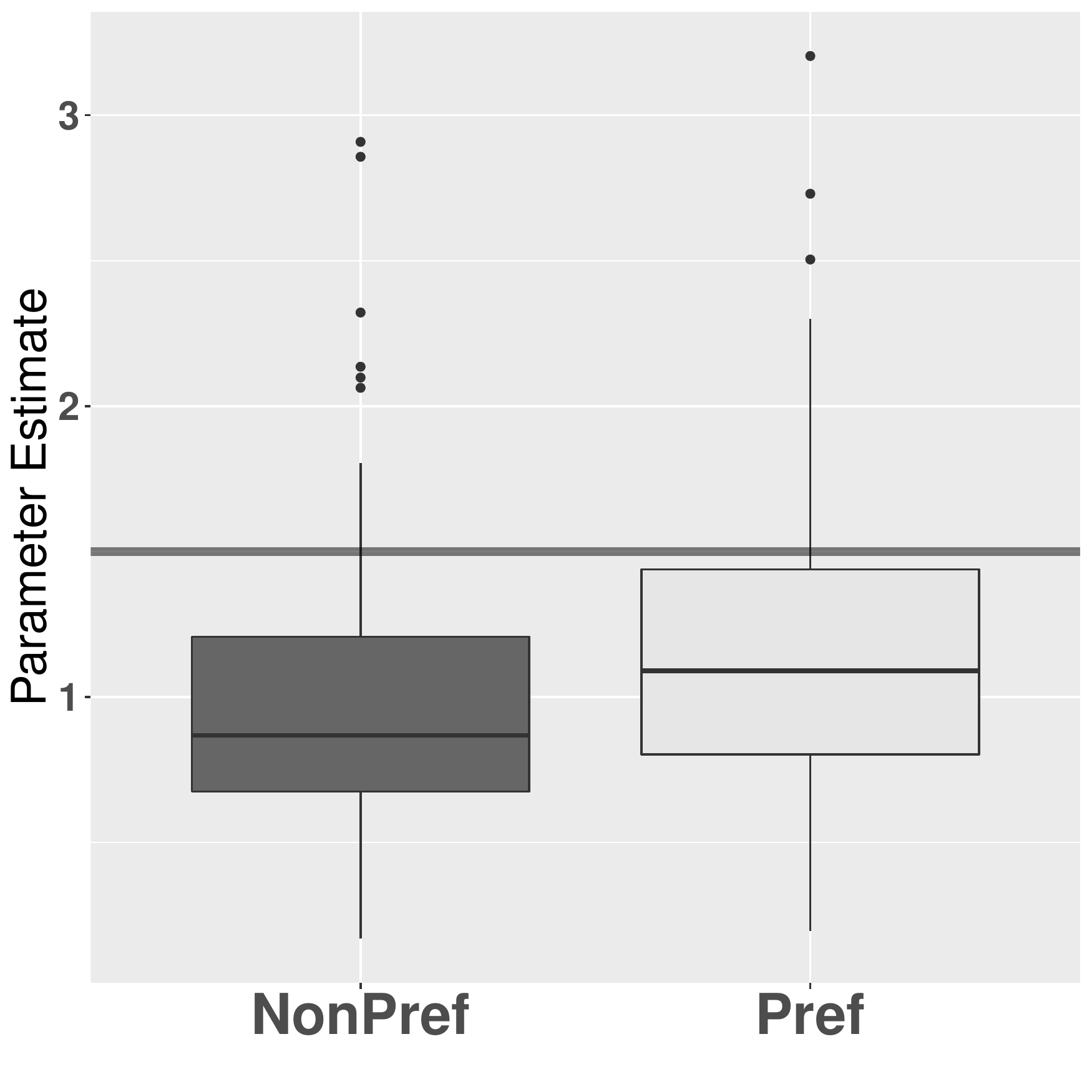}} 
\caption{Field parameter estimates over 100 preferentially sampled simulated data sets with true parameter values marked as grey lines. The abbreviations \texttt{NonPref} and \texttt{Pref} stand for the standard MLE (non-preferential) estimation and the one using the 
preferential Preferential-CRW model of Section \ref{sec:CTCRWmarinemammals}.}
\label{fig:hybridsimfield}
\end{figure}

Figure~\ref{fig:hybridsimfield} shows the boxplots of the 100 estimated
parameters for the spatial field process ($\btheta_F$) using each of the two likelihoods (standard
and accounting for preferential sampling). The grey horizontal lines represent
the true values.  As expected, when the model does not account for preferential
movement, the prevalence of lower temperatures in the sample introduces a
negative bias to the estimates for the mean parameter ($\mu$).  A similar
pattern is observed also for the scale ($\phi$) and marginal variance
($\sigma^2$).  The bias in the variance estimates is likely due to the tracks
avoiding higher temperature regions and recording temperatures with a reduced
range than they would otherwise.  This may also explain the negative bias in
the scale estimates.  In contrast, using the
Preferential-CRW model with \texttt{TMB} results in 
better parameter estimates. This is particularly noticeable for the estimates of
$\mu$. Movement parameter estimates are discussed in Appendix~\ref{appendix:movementestimates}.
%

%

\subsection{Prediction}\label{sec:simpred}
We now turn our attention to the predictions for the underlying spatial field $S$. 
The first set of predictions are computed via 
kriging with parameter estimates obtained from the standard model
that conditions on the observed locations, while the
preferential sampling ones correspond to the 
estimated mode of $[\bS|\bX,\bY;\btheta_{\text{opt}}]$, where $\btheta_{\text{opt}}$ are the parameter estimates from our preferential likelihood (see equation~\eqref{eq:TMBpointpred}). 

Predictions were computed on a $26\times 26$ lattice for each of the $M=100$ data sets
and we used two different measures of their quality. The root mean square prediction 
error (RMSPE) over the discrete domain is given by 
\begin{equation}\label{eq:biasandMSPE}
\operatorname{RMSPE}_i = \frac{1}{M}\sum_{j=1}^{M}\sqrt{\left(\bS_{j,i}-\hat{\bS}_{j,i}\right)^2},
\end{equation}
for each location $i=1, \ldots, N = 26^2$ on the prediction grid. Here 
$\bS_{j,i}$ is the true value of field for the $j$-th simulated data set $\bS_j$ at the $i$th prediction location ($\bS_j(\bx_i)$) and $\hat{\bS}_{j,i}$ is the corresponding predicted value. 

To compare the resulting predictions whilst accounting for 
their variances, we used Ignorance Scores~\citep{Roulston2002},
which are given by $\operatorname{IGN}(x)=-\log(p(x))$, 
where $p$ is the predictive density and $x$ is the target
 forecast which would be the true SST at that location~\citep{Siegert2014,Gneiting2007}. 
For each of the $j=1,\ldots, M= 100$ preferential simulations we calculated the 
Mean Ignorance Score (MIGN) of our latent field predictions as
\begin{equation}\label{eq:MIGN}
\operatorname{MIGN}_j=\frac{1}{N}\sum_{i=1}^{N}\left\{\frac{(\bS_{j,i} - 
\hat{\bS}_{j,i})^2}{2\hat{\sigma}_{j,i}^2}+ \log\hat{\sigma}_{j,i}\right\} \, 
\quad j = 1, \ldots, M \, , 
\end{equation} 
where $\hat{\sigma}_{j,i}^2$ is the prediction variance of $\hat{\bS}_{j,i}$. This measure gives an indication of model performance for each simulation 
averaged over the entire domain.
We also calculated location-specific Ignorance Scores, averaging the IGN of each location over the 100 
samples, we call them Location Ignorance Scores (LIGN): 
\begin{equation}\label{eq:LocationMIGN}
\operatorname{LIGN}_i=\frac{1}{M}\sum_{j=1}^{M}\left\{\frac{(\bS_{j,i} - 
\hat{\bS}_{j,i})^2}{2\hat{\sigma}_{j,i}^2}+ \log\hat{\sigma}_{j,i}\right\} \, 
\quad i = 1, \ldots, N \, . 
\end{equation} 
The LIGN gives an assessment of model prediction across each region of the domain.

To compare the ``standard'' predictions with the ``preferential sampling'' ones
we computed the corresponding differences of the above 3 measures:
\begin{align}\label{eq:plotmeasures}
\begin{split}
\operatorname{RMSPE}^{\text{Diff}}_i &= \operatorname{RMSPE}^{\text{P}}_i - \operatorname{RMSPE}^{\text{NP}}_i, \\
\operatorname{MIGN}^{\text{Diff}}_j &= \operatorname{MIGN}^{\text{P}}_j - \operatorname{MIGN}^{\text{NP}}_j, \\
\operatorname{LIGN}^{\text{Diff}}_i &= \operatorname{LIGN}^{\text{P}}_i - \operatorname{LIGN}^{\text{NP}}_i,
\end{split}
\end{align}
where $\text{NP}$ and $\text{P}$ indicate the values of the scoring functions for the 
non-preferential (standard) and preferential models, respectively. 

\begin{figure}
\centering     
\subfigure[$\operatorname{RMSPE}^{\text{Diff}}_i$]{\includegraphics[width=2.5in]{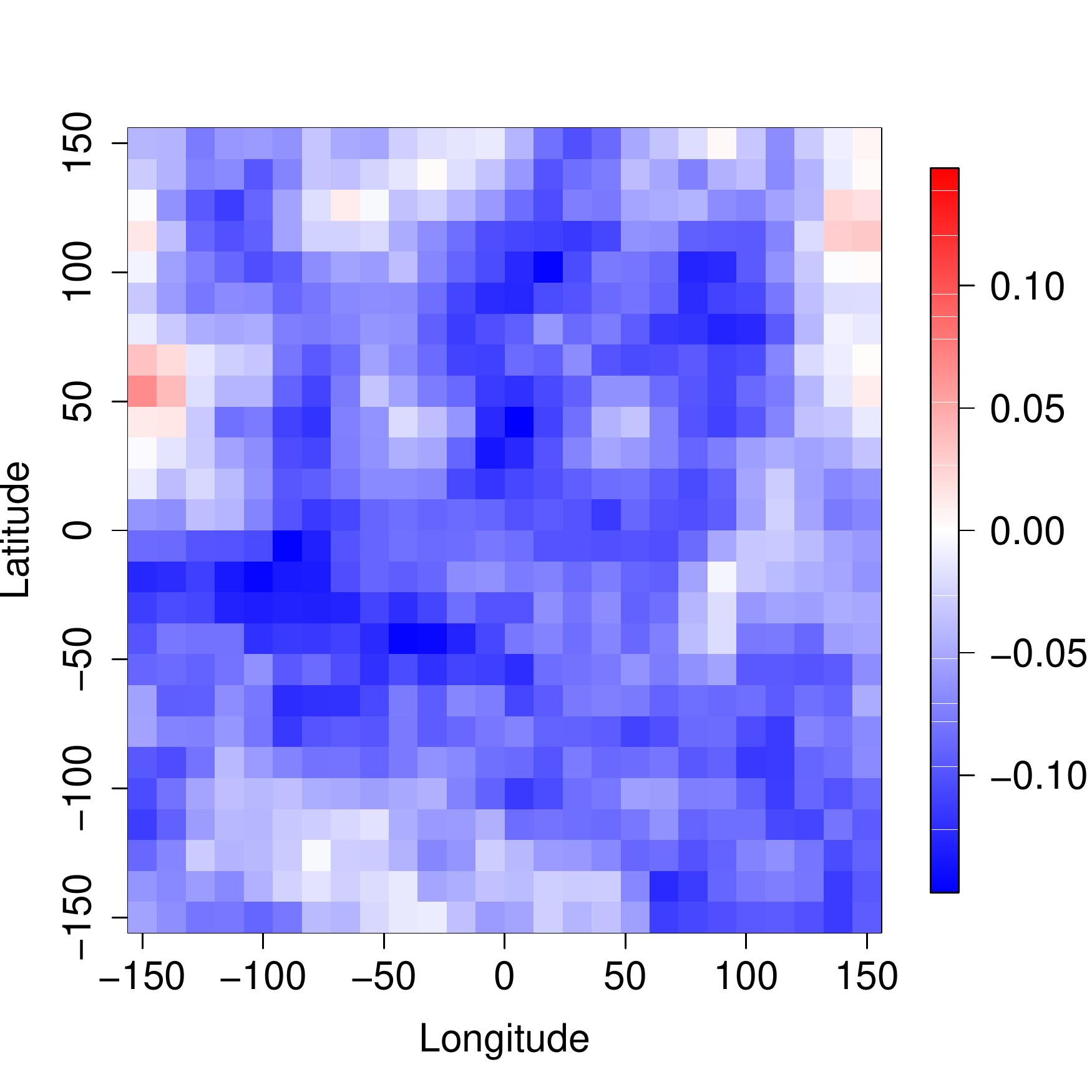}}
\subfigure[$\operatorname{MIGN}^{\text{Diff}}_j$]{\includegraphics[width=2.2in]{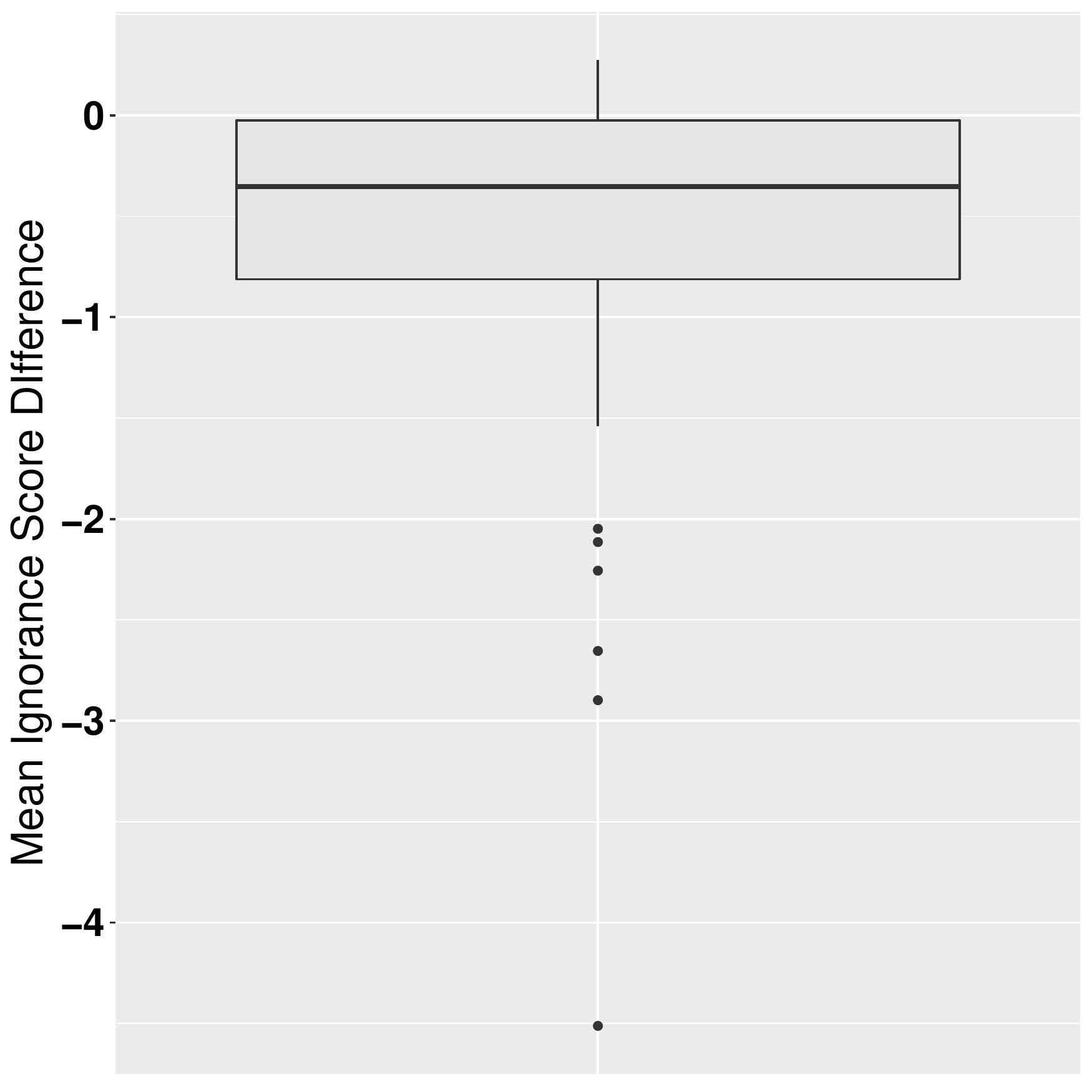}}
\subfigure[$\operatorname{LIGN}^{\text{Diff}}_i$]{\includegraphics[width=2.5in]{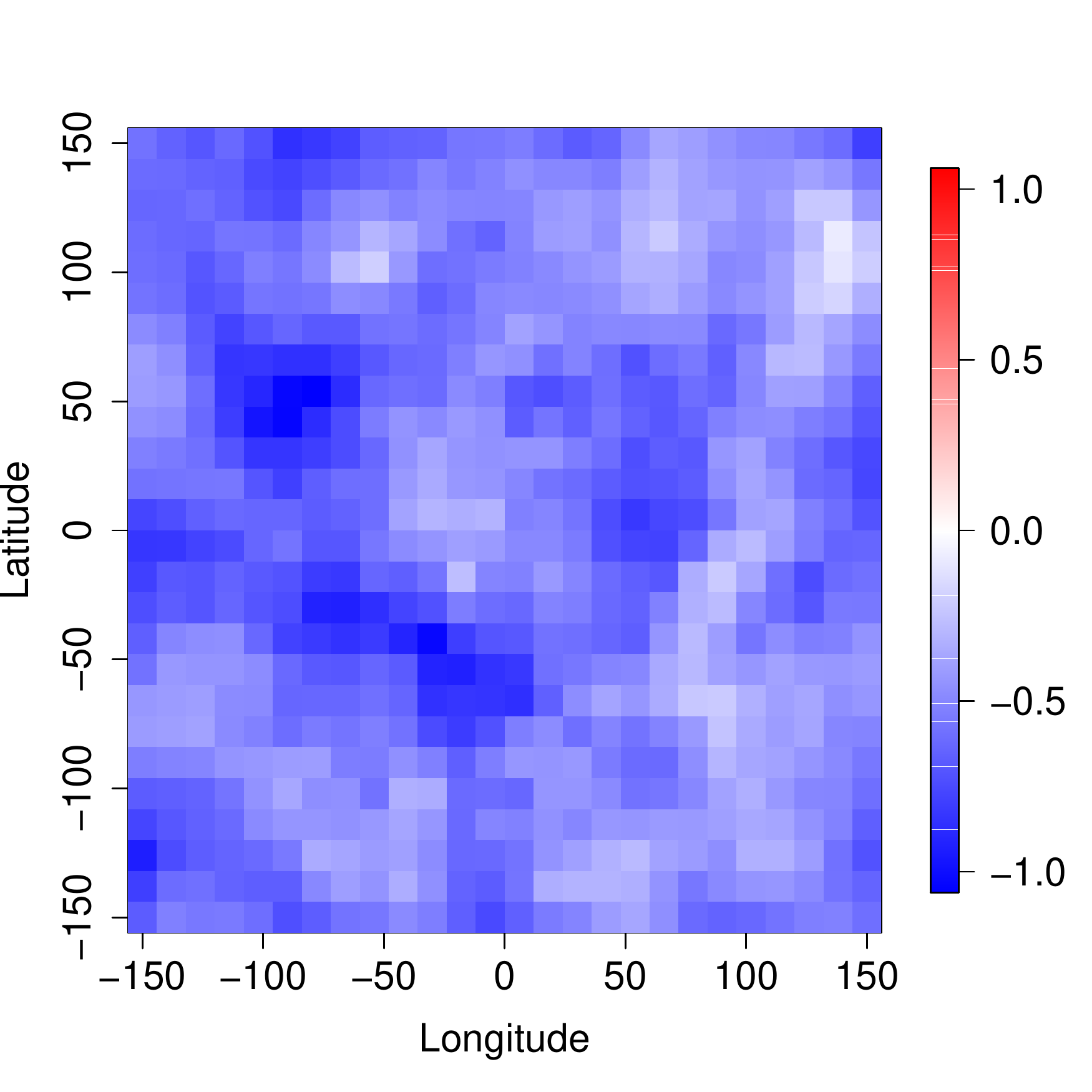}}
\caption{Comparison of Root Mean Squared Prediction Error (RMSPE) difference, Mean Ignorance Score (MIGN) difference and Location Ignorance Score (LIGN) difference respectively across 100 
preferentially sampled simulated data sets.}
\label{fig:hybridPrediction}
\end{figure}

The first panel in Figure~\ref{fig:hybridPrediction} shows the values of $\operatorname{RMSPE}^{\text{Diff}}_i$ 
at each location, colouring the areas in blue for which this measure was negative, 
which correspond to 
regions where the 
RMSPE's for preferential sampling predictions were better. 
Nearly all regions were predicted more accurately on average using the preferential sampling model, 
with only a small number of positive (red) locations.

The boxplot of the $M = 100$ differences in MIGN is displayed in the
second panel of Figure~\ref{fig:hybridPrediction}. Negative values of this difference imply 
that the $\text{P}$ model had a lower average Ignorance Score 
(over the region) for that particular simulation run. Again we see that the
preferential sampling model performs better than standard methods across the 
majority of simulated data sets. This conclusion is further supported by the third panel 
in Figure~\ref{fig:hybridPrediction} 
which shows $\operatorname{LIGN}^{\text{Diff}}_i$ for each lattice prediction location. 
All locations had a smaller Location Ignorance Score on average when using the preferential
sampling model,  
which interestingly includes the locations where point predictions were slightly inferior. 
This suggests that, even in this case, the prediction 
variances were more reasonable. 

\section{Real Data Example}\label{sec:realdata}
%
In this section we analyse Southern Indian ocean temperatures at a depth of 6 meters, which we will be calling sea surface temperatures (SSTs), using CTD (Conductivity-Temperature-Depth) sensor data from tags attached to Elephant Seals. These data were collected and made freely available as part of the MEOP (Marine Mammals Exploring the Oceans Pole to Pole) database~\citep{Roquet2013,Roquet2014} and is described in Section~\ref{sec:intro} and shown in Figure~\ref{fig:CoriolisEastFull}. 

Since these data are relatively near to the South Pole, there is a large change in true distance between latitudinal and longitudinal degrees over the domain. For example, we can see in Figure~\ref{fig:CoriolisEastFull} that the change in latitude is not constant in distance between $-45^\circ$ and  $-65^\circ$. This is due to the difficulties of representing locations from a 3-dimensional sphere on a 2-dimensional surface, which is of particular importance near the poles. One option is to use the Haversine formula to calculate the great-circle distance between two points in the latitude/longitude space~\citep{Robusto1957}. However, as is discussed by \cite{Jeong2015, Gneiting2013} among others, using the Mat\'{e}rn covariance along with great circle distances requires the use of a non mean-square differentiable processes ($0<\kappa\leq 0.5$), since the Mat\'{e}rn class is not isotropic otherwise.

\begin{figure}
\centering     
\includegraphics[width=3.25in]{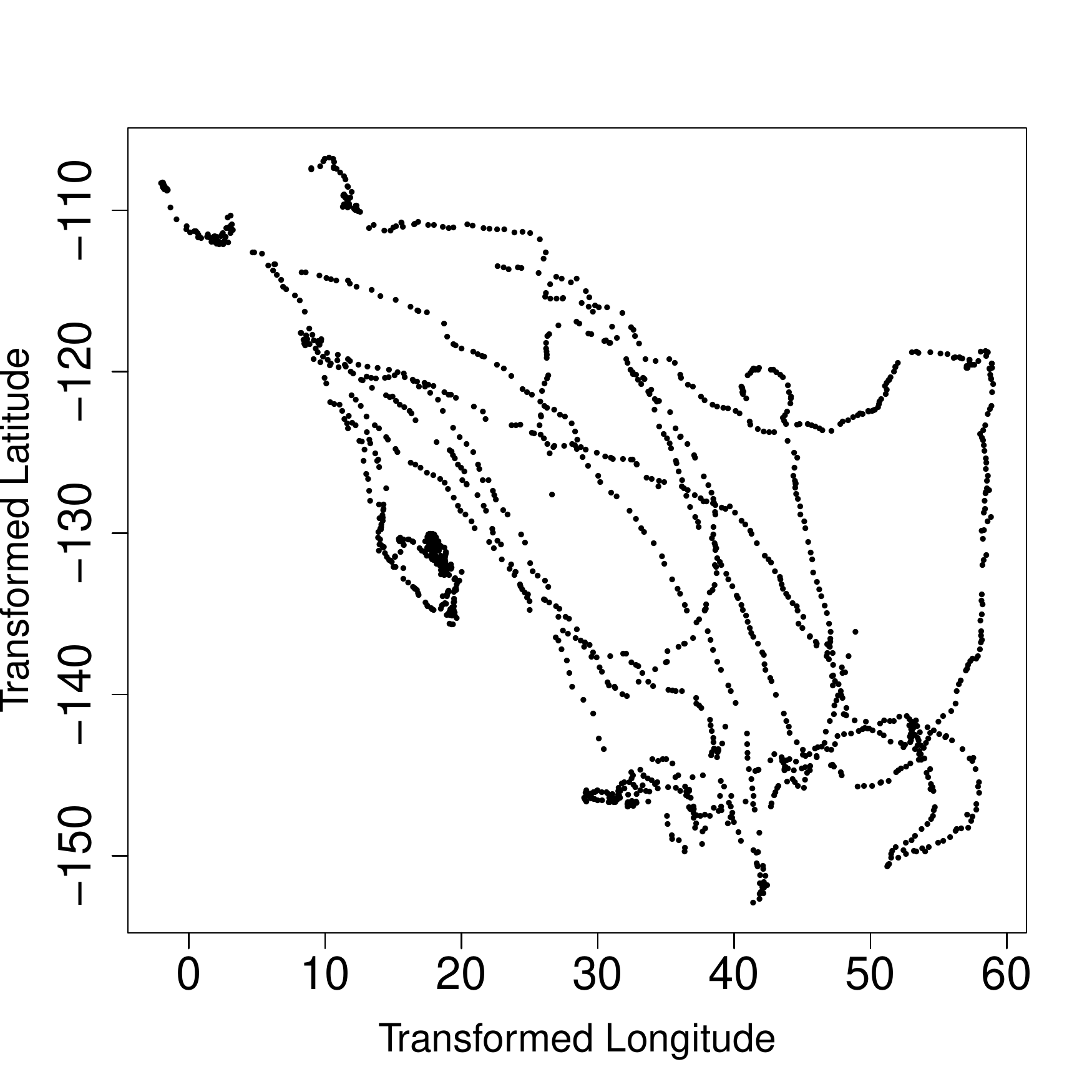}
\caption{The corresponding transformed locations from Figure~\ref{fig:CoriolisEastFull} using a scaled version of the Universal Transverse Mercator (UTM) projection (zone 43).}
\label{fig:roquetsubset}
\end{figure}

To more accurately represent distance over our domain but remaining in the Euclidean space, we transformed the sampling locations. Our transformation used a scaled version of the Universal Transverse Mercator (UTM) projection (zone 43) which can be seen in Figure~\ref{fig:roquetsubset}. This transformation provides us with a more accurate representation of distance than using raw latitudes and longitudes, whilst remaining in the 2-dimensional Euclidean space and retaining the isotropic properties of our correlation model. The values of our scale have no real-world interpretation, other than to provide more realistic scaled distances between sampled locations.

\begin{figure}
\centering     
\subfigure[Mean ($\mu$)]{\includegraphics[width=50mm]{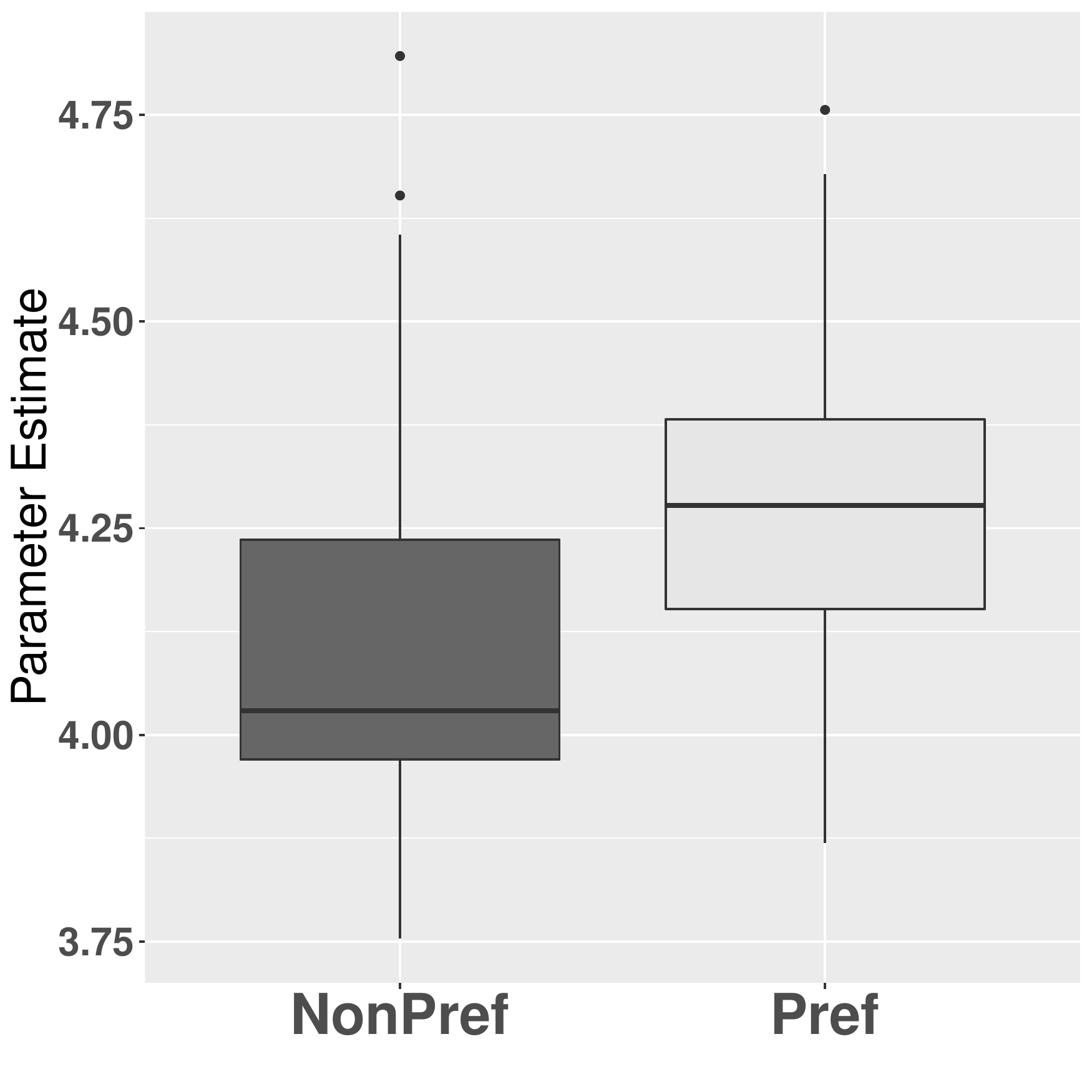}}
\subfigure[Scale ($\phi$)]{\includegraphics[width=50mm]{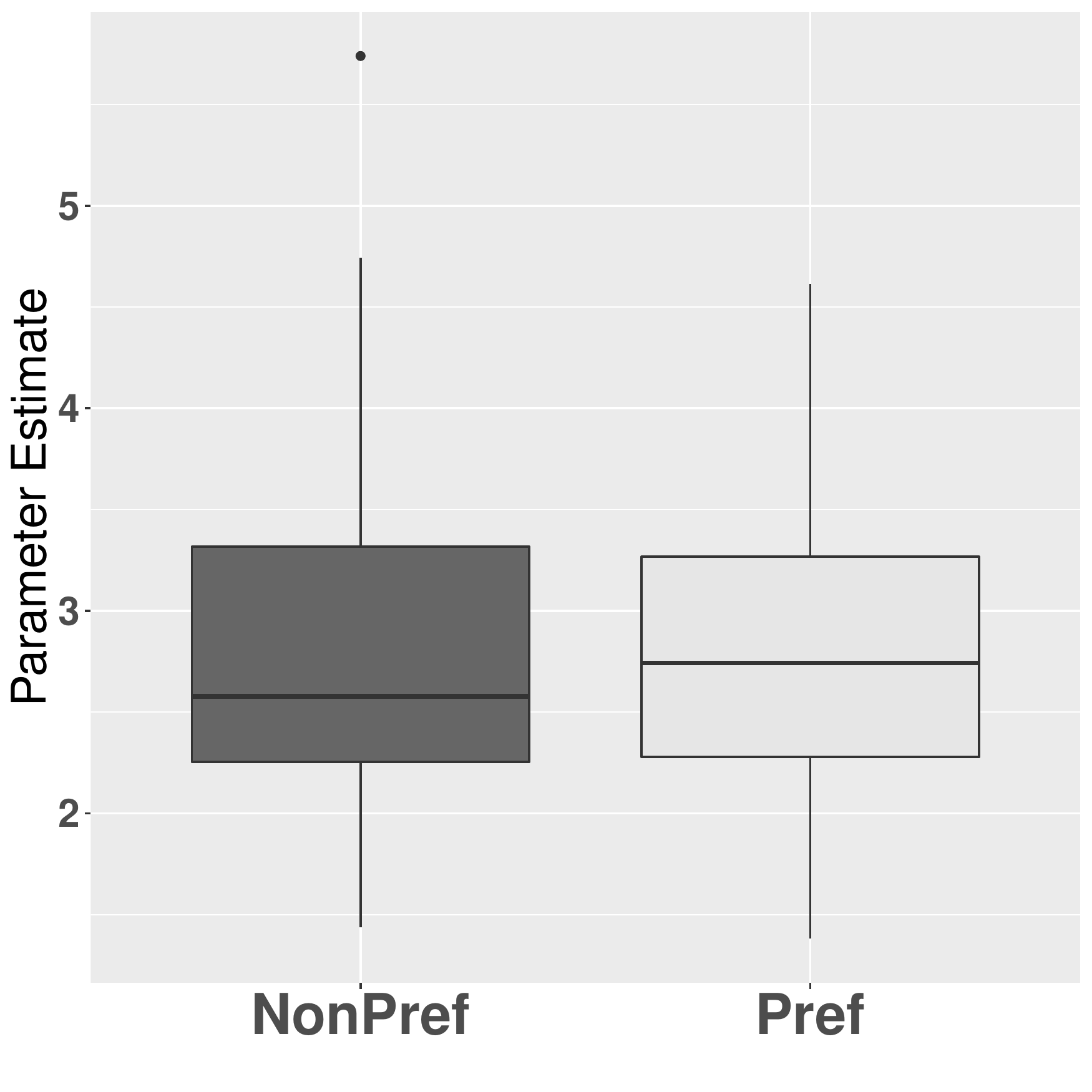}}
\subfigure[Variance ($\sigma^2$)]{\includegraphics[width=50mm]{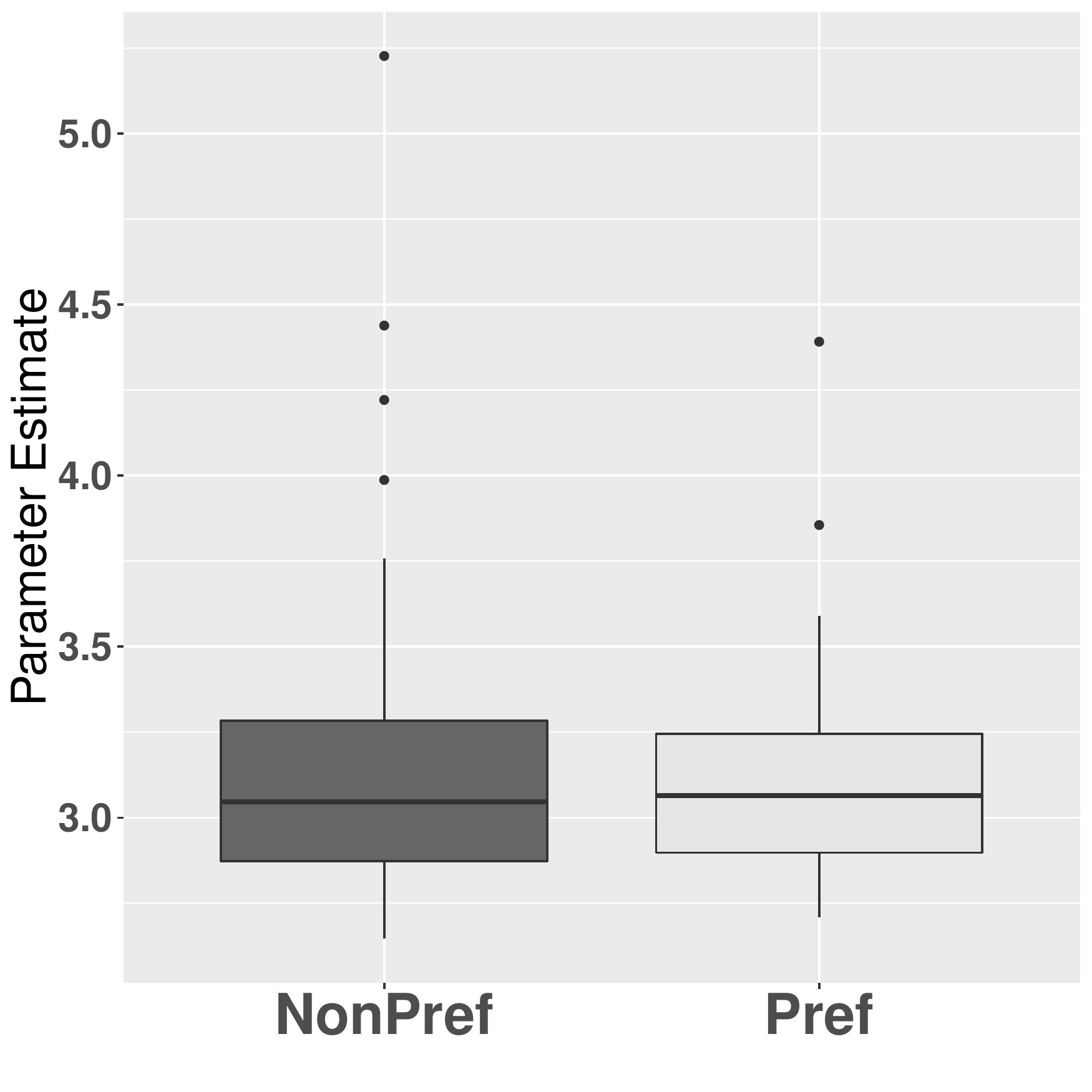}} 
\subfigure[$\alpha$]{\includegraphics[width=50mm]{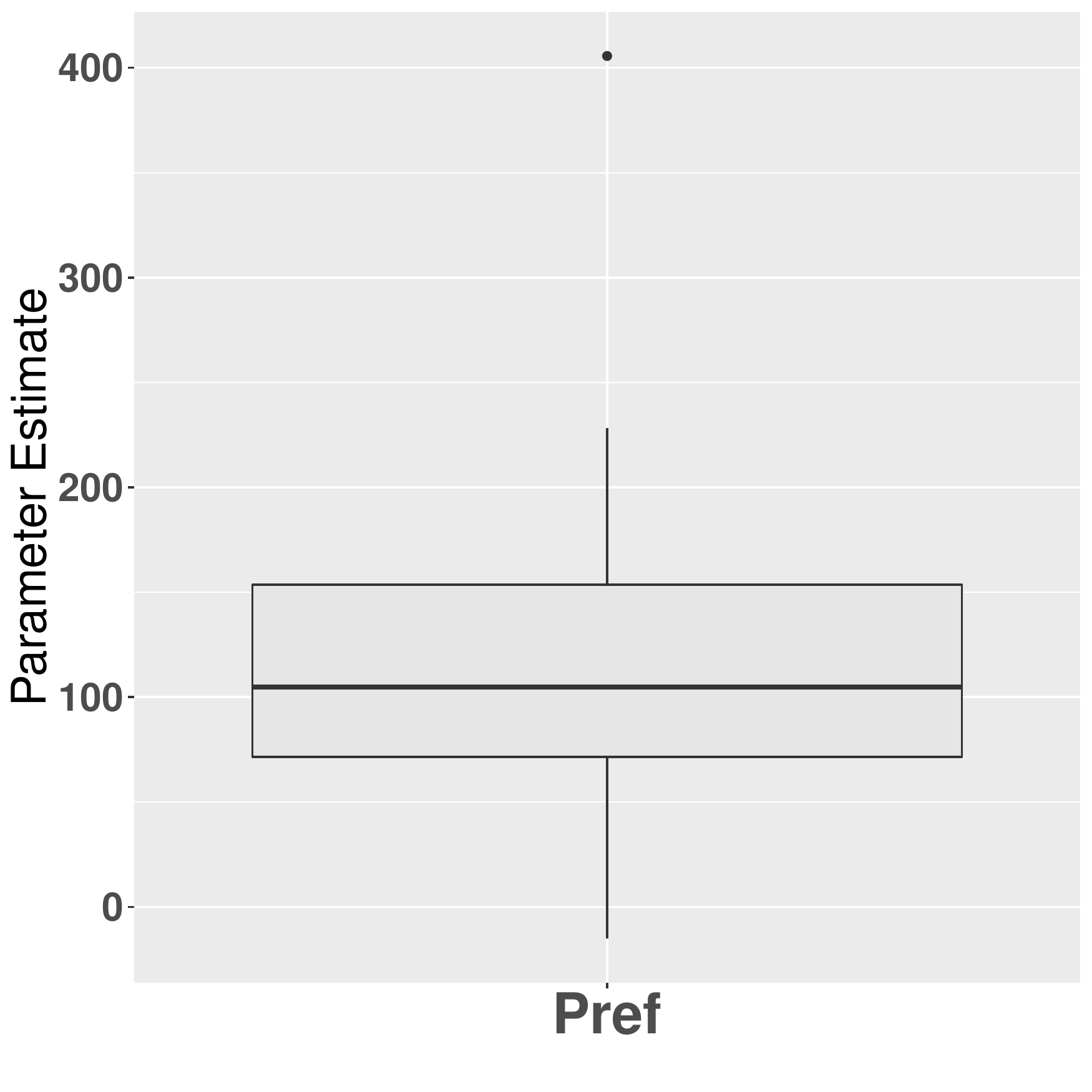}}
\caption{Comparison between preferential and standard MLEs for the preferential and field parameters over 50 data sets consisting of a sub-sample of 9 tracks with 40 observations each.}
\label{fig:fieldcorioliseast}
\end{figure}

We compare both the parameter estimates and corresponding field predictions using the PCRW model described in Section~\ref{sec:CTCRWmarinemammals}, with those obtained from a standard geostatistical model in which the sampling locations were considered independent of the temperature field. In order to control the computational complexity of the analysis and also to 
explore the sampling distribution of the field parameter estimators, we use 50 subsamples from the data. 
For each of these 50 replications we
randomly sampled 40 observations from each of the 9 tracks and estimated the parameters on the sub-sampled data. The resulting parameter estimates are displayed in Figure~\ref{fig:fieldcorioliseast}. 
%
We note that the difference between the two sets of estimates is very small in the scale and variance parameter estimates. However, there was an increase in the mean parameter estimate using the PCRW model. The consistently positive $\hat{\alpha}$ estimates suggest that there was a tendency of the animals to move towards cooler regions according to our model, which explains the increased mean parameter estimates. 

For predictive assessment of the ocean temperatures obtained from marine mammal tags we compare our model predictions with monthly average temperature fields obtained via the Simple Ocean Data Assimilation ocean/sea ice reanalysis (SODA)~\citep{Carton2008}, specifically the SODA version 3 (SODA3) reanalysis~\citep{Carton2018}, which uses all temperature and salinity profiles from the World Ocean Database. This data is available for depths between 5 and 5000 meters below the surface at a spatial resolution of 0.5 degrees latitude and longitude. Data is available from January 1980 to present. 

\begin{figure}
\centering     
\subfigure[SODA3 Temperature Field]{\includegraphics[width=2.4in]{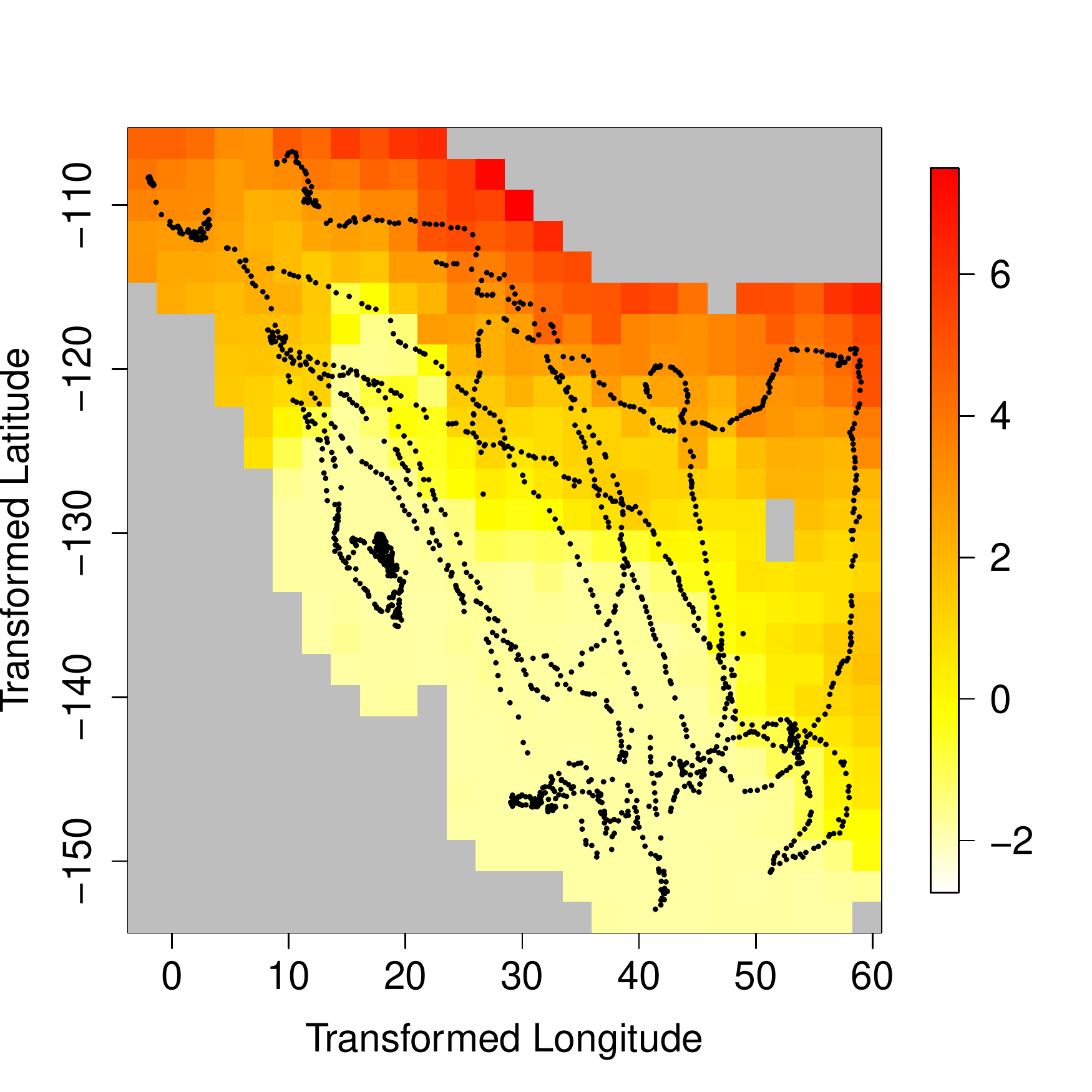}}
\subfigure[Median of Prediction Difference]{\includegraphics[width=2.4in]{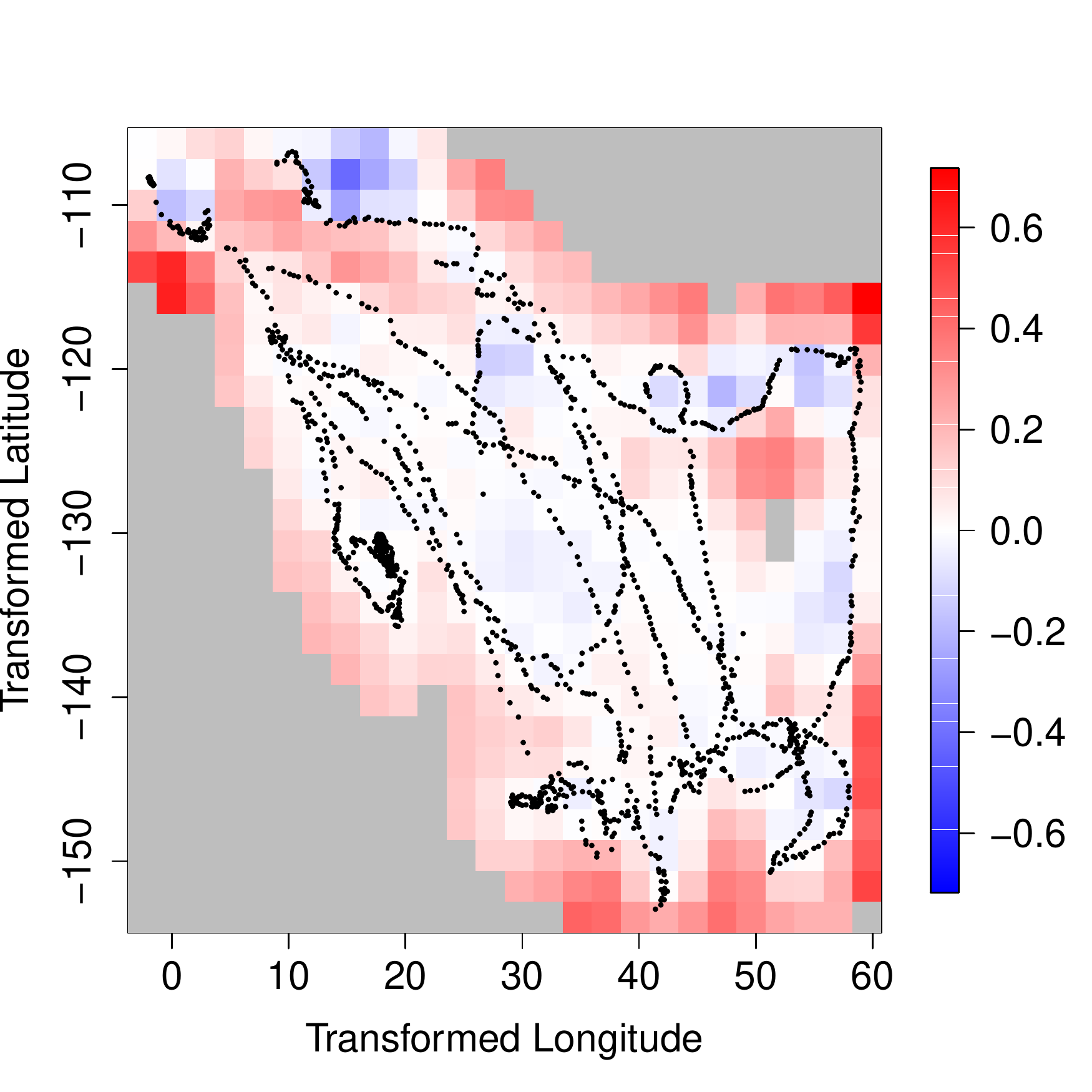}}
\caption{Panel (a): Monthly average field obtained via the Simple Ocean Data Assimilation ocean/sea ice reanalysis version 3 (SODA3) analysis~\citep{Carton2018}. Panel (b): Median of prediction difference between preferential and non-preferential models. Positive values imply that the preferential model tends to increases temperature prediction at that location and negative the opposite. 
}
\label{fig:RoquetPredComp}
\end{figure}

The first panel of Figure~\ref{fig:RoquetPredComp} shows the SODA3
monthly average ocean temperature field for August for a depth of 5 meters. 
We compare our predictions with these monthly averages at $N=461$ locations which were chosen as those 
on the original $26\times 26$ lattice points which were close enough to sampling locations to obtain kriging predictions that did not simply tend to the mean trend parameter $\hat{\mu}$. The locations on the lattice which were not used are shown in grey. 
To compare point predictions we consider the quantiles of the difference between the two prediction methods on each of the data sets at each prediction location. In other words we consider the quantiles of each coordinate of the vector $\bD=(\bold{\textit{D}}_1, \ldots, \bold{\textit{D}}_N)$ such that
\begin{equation}\label{eq:quantilediff}
\bold{\textit{D}}_i = (\hat{S}^\textit{P}_{1,i}-\hat{S}^\textit{NP}_{1,i}, \ldots,
\hat{S}^\textit{P}_{50,i}-\hat{S}^\textit{NP}_{50,i})^{T},
\end{equation}
for $i=1,\ldots,461$, where $\hat{S}^\textit{P}_{j,i}$ is the prediction at location $i$ in simulation $j$ using the preferential model and $\hat{S}^\textit{NP}_{j,i}$ the equivalent using the non-preferential model.
We plot the $50\%$ quantile of $\bD$ in the second panel of Figure~\ref{fig:RoquetPredComp}. As we would expect with the positive $\alpha$ estimates, most areas further from sampling locations had an increased SST prediction using the PCRW model. Interestingly however, the PCRW model actually tended to decrease SST prediction in certain areas of the West and North-East predictive region. We discuss the magnitude of these prediction differences further in Appendix~\ref{appendix:predif}.

\begin{figure}
\centering     
\subfigure[$\operatorname{RMSPE}^{\text{Diff}}_i$]{\includegraphics[width=2.4in]{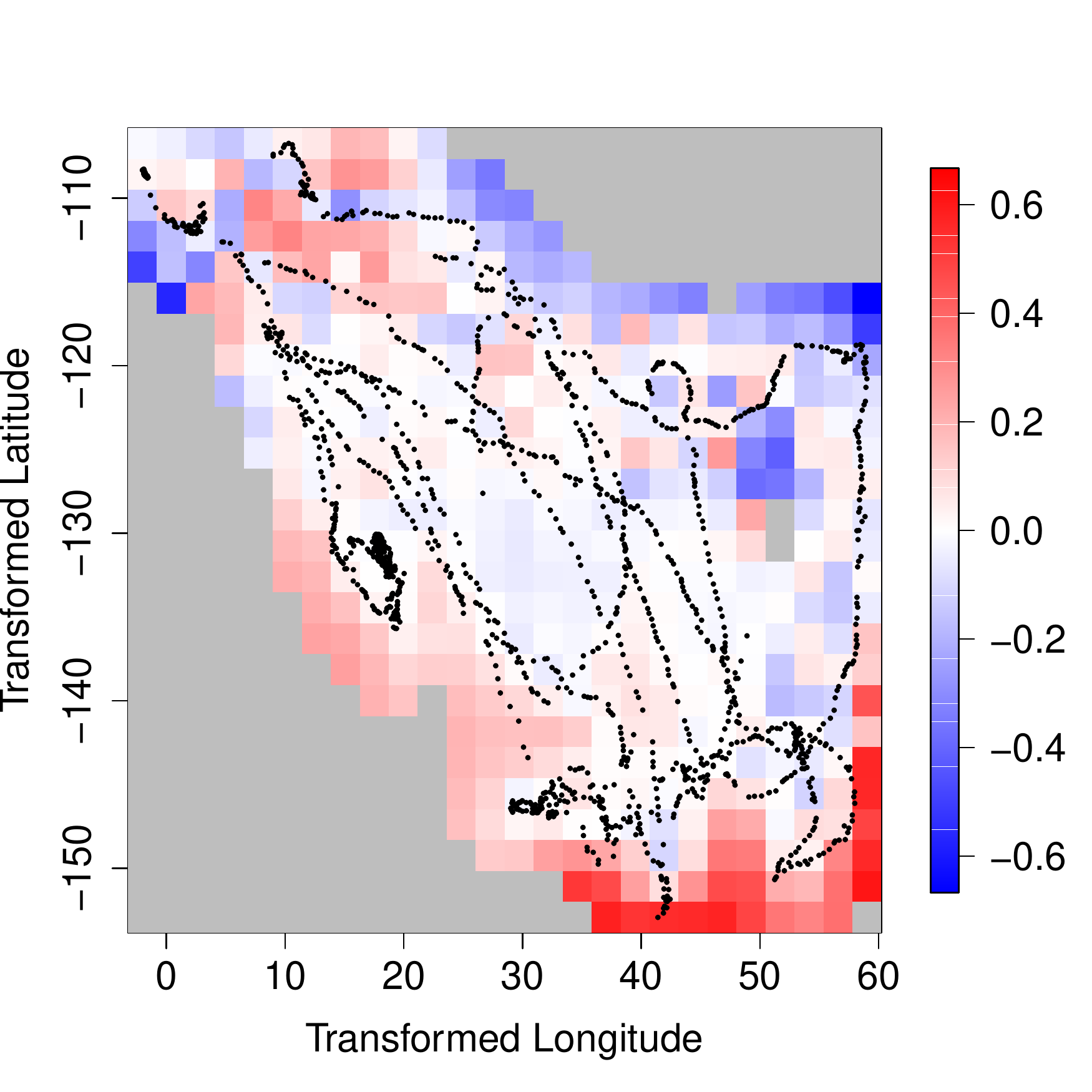}}
\subfigure[$\operatorname{MIGN}^{\text{Diff}}_j$]{\includegraphics[width=2.1in]{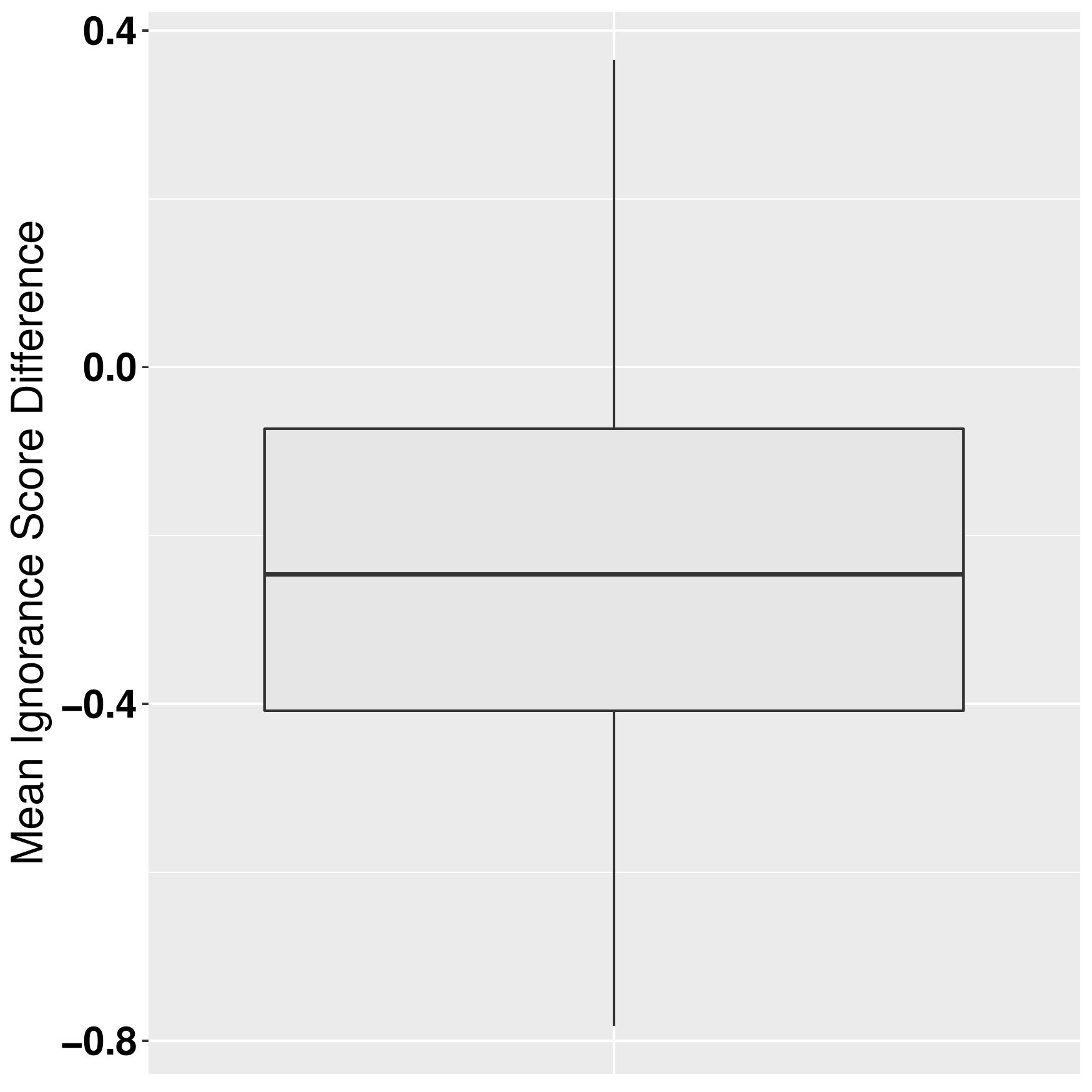}}
\subfigure[$\operatorname{LIGN}^{\text{Diff}}_i$]{\includegraphics[width=2.4in]{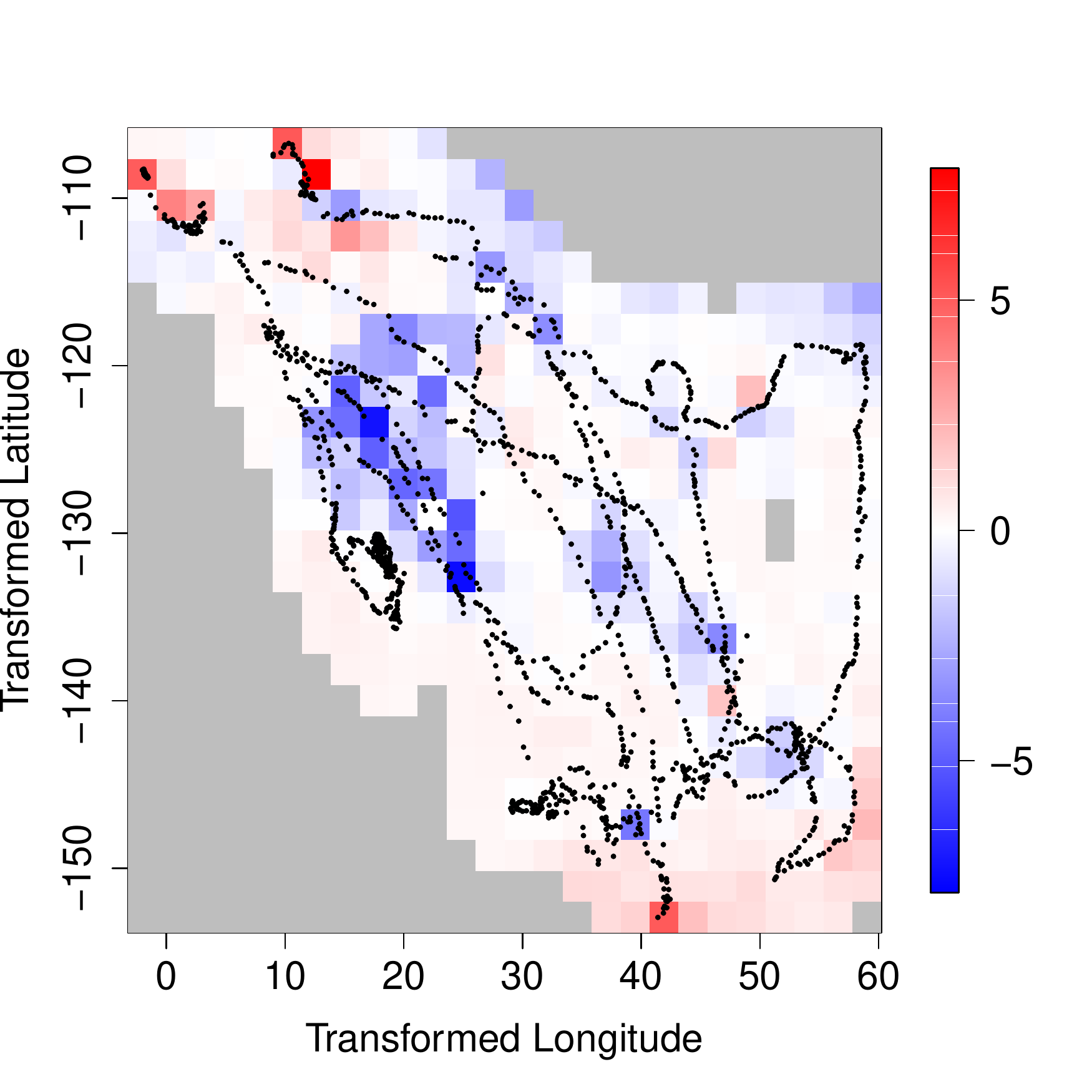}}
\caption{Comparison of Root Mean Square Prediction Error (RMSPE) difference, mean ignorance score (MIGN) difference and location ignorance score (LIGN) difference respectively across 50 simulated data sets between preferential and non-preferential temperature predictions.}
\label{fig:RoquetPrediction}
\end{figure}

To assess the prediction accuracy compared to the SODA3 data set we consider ignorance scores and RMSPE's in the same manner as in Section~\ref{sec:simpred}. We plot the difference in RMSPE, MIGN and LIGN defined in \eqref{eq:plotmeasures} in Figure~\ref{fig:RoquetPrediction}. 
The first panel shows that the PCRW model tended to reduce RMSPE in comparison to non-preferential prediction in the Northern regions, whilst underperforming in areas of the South. However, the LIGN in the third panel shows improved ignorance scores in general using the preferential model. An indication of superior prediction performance is through the second panel of Figure~\ref{fig:RoquetPrediction} which shows the majority of MIGN's across the 50 subsamples
were negative, and indication of better predictions using the PCRW model.

To summarise, the PCRW model applied using \texttt{TMB} showed differences in mean parameter estimation when compared to the naive method and identified possible tendency of sampler movement towards the cooler regions. This translated to many areas of increased SST prediction when using the preferential model in spatial prediction, but not across the entire spatial region. Small changes in SST prediction as shown by our model may be of increased importance when dealing with complex systems as ocean temperatures. Although this simulation used a simplistic relationship between the sampler and the process being modelled, our results suggest that developing more realistic models for this type of data can improve the resulting statistical inference about our ocean climates.

\section{Discussion} \label{sec:disc}
We have shown the damaging effect that preferential sampling may have on statistical inferences based on spatial models where monitoring locations are not stationary. The simulation experiments reported in Section~\ref{sec:simulateddata} illustrate how predictions may be improved when accounting for the preferential nature of movement in the sampling model. This is evident in the parameter estimates and also in the predictive performance. Here the combination of corrected parameters and a predictive distribution that accounts for the relationship between the sampling locations and the spatial field of interest improves notably upon the simple extrapolation of kriging. When we compare the results of the standard method (that conditions on the observed locations) and the Preferential-CRW model in our motivating example we observe increased mean parameter estimates for the SST field, whilst the predicted SST fields show consistent differences.

These analyses highlight the importance of expanding preferential sampling methods to the case of non-stationary sampling locations, which is becoming a more prevalent situation. We show how implementing a Laplace approximation to the likelihood function via the \texttt{R} package \texttt{TMB} allows for flexible movement model specification. This method can expand beyond the typical point process models to sampling processes derived from movement models which may depend on our field of interest $S$. This is just the first step in incorporating preferential sampling into the statistical analyses of tagged marine mammals. Considering the observed measurement locations to contain some measurement error by assuming there is an latent true location state would help to account for the uncertainty in the Argos location estimates, whilst there are many other applications such as land animals and other non-stationary sampling processes as the next step in expanding our applications and methodology. 
Furthermore, a natural extension of this research is to consider data at any depth. Although SST analysis is a useful first step, an important application of these methods would be to aid in mapping the water masses at depths unobservable by satellites. Theoretically little would change in comparison to the SST data we consider here in analyses at specific depths, however the location accuracy underwater is reduced. In this case, location estimates could inferred based on accelerometer data using methods such as the dead reckoning algorithm~\citep{Wilson1988, Liu2015} and ocean temperature at specific depths analysed to build a water temperature profile.

\appendix
\section{Using Smooth Random Fields with TMB}\label{appendix:smoother}
Many geostatistical models proposed to analyse spatial data assume that the underlying
random field is continuous (but not necessarily differentiable). In what follows we show how \texttt{TMB} can be used to efficiently employ mean-square differentiable Gaussian fields (GFs) in our models, using the approach by \cite{Lindgren2011} who show how Gaussian Markov random field (GMRF) representations of GFs with Mat\'{e}rn covariance (as defined in \eqref{eq:matern0}), can be constructed through the solution to a stochastic partial differential equation (SPDE) when driven by white noise.

Methods that utilise GMRF representations of GFs require constructing a sparse precision matrix $\bQ$ that closely represents the covariance of the GF. We follow directly after (9) in \cite{Lindgren2011} but note  the change in notation from $\kappa$ in the original work to $\phi$ here and that in our 2-dimensional application $\alpha=\kappa + 1$ where $\kappa$ is the smoothness parameter in \eqref{eq:matern0}.

\cite{Lindgren2011} write the precision matrix $\bQ_{\alpha,\phi}$ as a combination of $m\times m$ matrices $\bC, \bG$ and $\bK_\phi$,
\begin{align*}
\bC_{i,j} &= \langle\psi_i, \psi_j\rangle, \\
\bG_{i,j} &= \langle\Delta\psi_i, \Delta\psi_j\rangle, \\
(\bK_\phi)_{i,j} &= \phi^{-2}\bC_{i,j} + \bG_{i,j},
\end{align*}
where $m$ is the number of vertices in the triangulation of the domain.

This combination depends on $\alpha$ but can be calculated recursively as
\begin{align*}
\bQ_{1,\phi} &= \bK_\phi,\\
\bQ_{2,\phi} &= \bK_\phi\bC^{-1}\bK_\phi, \\
\bQ_{\alpha,\phi} &= \bK_\phi\bC^{-1}\bQ_{\alpha-2,\phi}\bC^{-1}\bK_\phi, \quad \text{for}\quad \alpha=3,4,\ldots.
\end{align*}
Unfortunately $\bC^{-1}$ is dense, but \cite{Lindgren2011} show that $\bC$ can be replaced by the diagonal matrix $\tilde{\bC}$, where $\tilde{C}_{i,i}=\langle\psi_i, 1\rangle$. Hence $\tilde{\bC}$ is sparse and the resulting precision matrix $\bQ_{\alpha,\phi}$ also sparse. Further details can be found in \cite{Lindgren2011, Simpson2012, Lindgren2015}.

\subsection{Implementation of $\kappa=2$ in TMB}
\texttt{R-INLA} only implements the SPDE approximation for $\alpha\in(0,2]$. At the largest smoothness of $\alpha=2$ ($\kappa=1$), the corresponding Mat\'{e}rn field is mean square continuous but not mean square differentiable. In some applications this might not suit the problem at hand (for example, when derivatives of the spatial field may be part of the model). Below we extend the current approximations to the case $\alpha=3$ ($\kappa=2$) and show how this can be implemented in \texttt{TMB} via the built-in compatibility with \texttt{R-INLA}.

To begin, note that for $\alpha=2$ we can expand the formula for $\bQ_{2,\phi}$ above as follows:
\begin{align*}
\bQ_{2,\phi} &= \bK_\phi\bC^{-1}\bK_\phi, \\
 &= (\phi^{-2}\bC + \bG)\bC^{-1}(\phi^{-2}\bC + \bG), \\
 &= \phi^{-4}\bC + 2\phi^{-2}\bG + \bG\bC^{-1}\bG, \\
 &= \phi^{-4}\bM_0 + 2\phi^{-2}\bM_1 + \bM_2 \, ,
\end{align*}
where $\bM_0=\bC,\bM_1=\bG$ and $\bM_2=\bG\bC^{-1}\bG$. Note that the matrices $\bM_1, \bM_2$ and $\bM_3$ do not depend on $\phi$ or $\kappa$ and can be computed with the R-INLA function \texttt{inla.spde2.matern}. 
Now we expand $\bQ_{3,\phi}$:
\begin{align*}
\bQ_{3,\phi} &= \bK_\phi\bC^{-1}\bQ_{1,\phi}\bC^{-1}\bK_\phi, \\
 &= \bK_\phi\bC^{-1}\bK_\phi\bC^{-1}\bK_\phi, \\
 &= (\phi^{-2}\bC + \bG)\bC^{-1}(\phi^{-2}\bC + \bG)\bC^{-1}(\phi^{-2}\bC + \bG), \\
 &= \phi^{-6}\bC + 3\phi^{-4}\bG + 3\phi^{-2}\bG\bC^{-1}\bG +\bG\bC^{-1}\bG\bC^{-1}\bG , \\
 &= \phi^{-6}\bM_0 + 3\phi^{-4}\bM_1 + 3\phi^{-2}\bM_2 +\bM_2\bM_0^{-1}\bM_1.
\end{align*}
Hence we can use $\bM_0,\bM_1$ and $\bM_2$ provided by \texttt{inla.spde2.matern} to construct $\bQ_{3,\phi}$ for a solution to the SPDE approximation for $\alpha=3$ ($\kappa=2$). We simply need to take a new combination of these matrices which in \texttt{TMB} can be done easily within the function that computes the likelihood. 
\section{Preferential-CRW Models with a Field-Dependent Velocity Term}\label{appendix:CTCRWdifficulties}
If one considers a velocity term $\bv$ defined in \eqref{eq:velocityapprox} that may depend on both the locations $\bX$ and the underlying field of interest $S$, the likelihood quickly becomes intractable. Considering a latent velocity state in the model, the full 
likelihood is
\begin{align}
\label{eq:motilityfactor}
\begin{split}
[\bX, \bY; \btheta] &= \int\int[\bX, \bY, \bS,  \bv; \btheta]\mathrm{d}\bS\mathrm{d}\bv, \\
		&=	\int\int[\bY|\bX, \bS, \bv; \btheta_F][\bX, \bS,  \bv; \btheta]\mathrm{d}\bS\mathrm{d}\bv, \\
		&=	\int\int[\bY|\bX, \bS,  \bv; \btheta_F][\bX| \bS,  \bv; \btheta_L][\bS,  \bv; \btheta]\mathrm{d}\bS\mathrm{d}\bv, \\
		&=	\int\int[\bY|\bX, \bS,  \bv; \btheta_F][\bX| \bS,  \bv; \btheta_L][\bv|\bS; \btheta_L][\bS; \btheta_F]\mathrm{d}\bS\mathrm{d}\bv \, .
\end{split}
\end{align}
Notice that this factorisation includes the term $[\bv|\bS; \btheta_L]$. If the velocity was to only depend on the latent field $\bS$ and not on the location $\bX$, then we could evaluate the likelihood. However, this makes little biological sense in practice. Preferential movement induced by a latent velocity would require this velocity to depend on the field $\bS$ at the current location $\bX$ of the animal. In other words the velocity would be dependent not just on the latent field but the current location within this field. However, this means that we would only have a tractable form for $[\bv|\bS, \bX; \btheta_L]$ and would therefore need to evaluate
\begin{align*}
[\bv|\bS; \btheta_L] &= \int [\bX, \bv|\bS; \btheta_L] \mathrm{d}\bX, \\
		&=  \int [\bv|\bS, \bX; \btheta_L] [\bX|\bS; \btheta_L] \mathrm{d}\bX \, .
\end{align*}
which is  difficult to compute due to the complexity of $[\bX|\bS; \btheta_L]$. The factorisation of the form in \eqref{eq:motilityfactor} appears intractable and therefore the Preferential-CRW model in which velocities depend on location and the latent field become difficult to implement using likelihood methods when preferential sampling may be present.
\section{Preferential CRW Likelihood}\label{appendix:newlik}
In Section~\ref{sec:prefTMB} we outlined how to evaluate the preferential model using a Laplace approximation to the likelihood function. In previous cases this factorisation was of the form $[\bX,\bY; \btheta]=\int[\bX,\bY,\bS; \btheta]\mathrm{d}\bS$, however in the case of the Preferential-CRW model we have a second latent vector in the behavioural states $\bbeta=(\beta_{t_{1}},\ldots,\beta_{t_{n}})$. Therefore we need to re-specify the full likelihood for our Laplace approximation routine.

\begin{align}
\label{eq:PCRWfactorisation}
\begin{split}
[\bX, \boldsymbol{Y}; \btheta] &= \int\int[\bX, \bY, \bS,  \bbeta; \btheta]\mathrm{d}\bS\mathrm{d}\bbeta, \\
		&=	\int\int[\bY|\bX, \bS,  \bbeta; \btheta_F][\bX, \bS,  \bbeta; \btheta]\mathrm{d}\bS\mathrm{d}\bbeta, \\
		&=	\int\int[\bY|\bX, \bS,  \bbeta; \btheta_F][\bX| \bS,  \bbeta; \btheta_L][\bS,  \bbeta; \btheta]\mathrm{d}\bS\mathrm{d}\bbeta, \\
		&=	\int\int[\bY|\bX, \bS,  \bbeta; \btheta_F][\bX| \bS,  \bbeta; \btheta_L][\bbeta|\bS; \btheta_L][\bS; \btheta_F]\mathrm{d}\bS\mathrm{d}\bbeta.
\end{split}
\end{align}
Notice that in the Preferential-CRW model we have $[\bbeta|\bS; \btheta_L]=[\bbeta; \btheta_{L2}]$ so therefore in the Laplace approximation implemented using \texttt{TMB}  we will need to redefine the joint  negative log-likelihood as 
\begin{equation*}
-\log([\bX, \bY, \bS, \bbeta; \btheta]) = -\log\left([\bY|\bX, \bS,  \bbeta; \btheta_F][\bX| \bS,  \bbeta; \btheta_L][\bbeta; \btheta_{L2}][\bS; \btheta_F]\right).
\end{equation*}
\section{Simulation Details}\label{appendix:simdetails}
\subsection{Data Generation} \label{appendix:datageneration}
To generate each track, we initialised a starting location chosen uniformly at random over the 2-dimensional domain $[-150, 150]\times [-150, 150]$, then simulated 360 observations with the time between consecutive observations following an exponential distribution with rate parameter $\lambda=10$. The first 60 positions of each animal were considered a burn-in period and discarded, resulting in 300 remaining observations. Finally the track was thinned by taking every 3rd observation, to retain a final track of 100 observations.

For each data set we simulated a random field $S$ over the domain and discretised it on a $51\times 51$ grid. We used a Mat\'{e}rn covariance function as in \eqref{eq:matern0} with smoothness parameter $\kappa=2$, scale $\phi=25$, marginal variance $\sigma^2=1.5$ and a constant mean $\mu=5$. Since we know the true field in generating the tracks, the gradient of the field used to direct movement was approximated using finite differences from points not necessarily on the grid. For the movement model we set $\sigma_{\beta}=0.1, \bSigma = 3 \, \bI_2$, where $\bI_2$ denotes the $2 \times 2$ identity matrix, $\alpha=100$ and initiated the behavioural states at $\beta_0=-1.5$.
From \eqref{eq:PCRWdrift} we see that this choice initialises primarily directed movement
($f(\beta_0)=0.18$), but with a slight influence of the foraging (preferential) function $\phi(\cdot)$. 
We assume that $\tau^2$ is a known parameter due to the assumption that the measuring device will have a known sampling error and as it is commonly done in the literature, we also assume that the smoothness parameter $\kappa$ is known.

\subsection{Movement Parameter Estimates}\label{appendix:movementestimates}
\begin{figure}
\centering     
\subfigure[$\alpha$]{\includegraphics[width=2.3in]{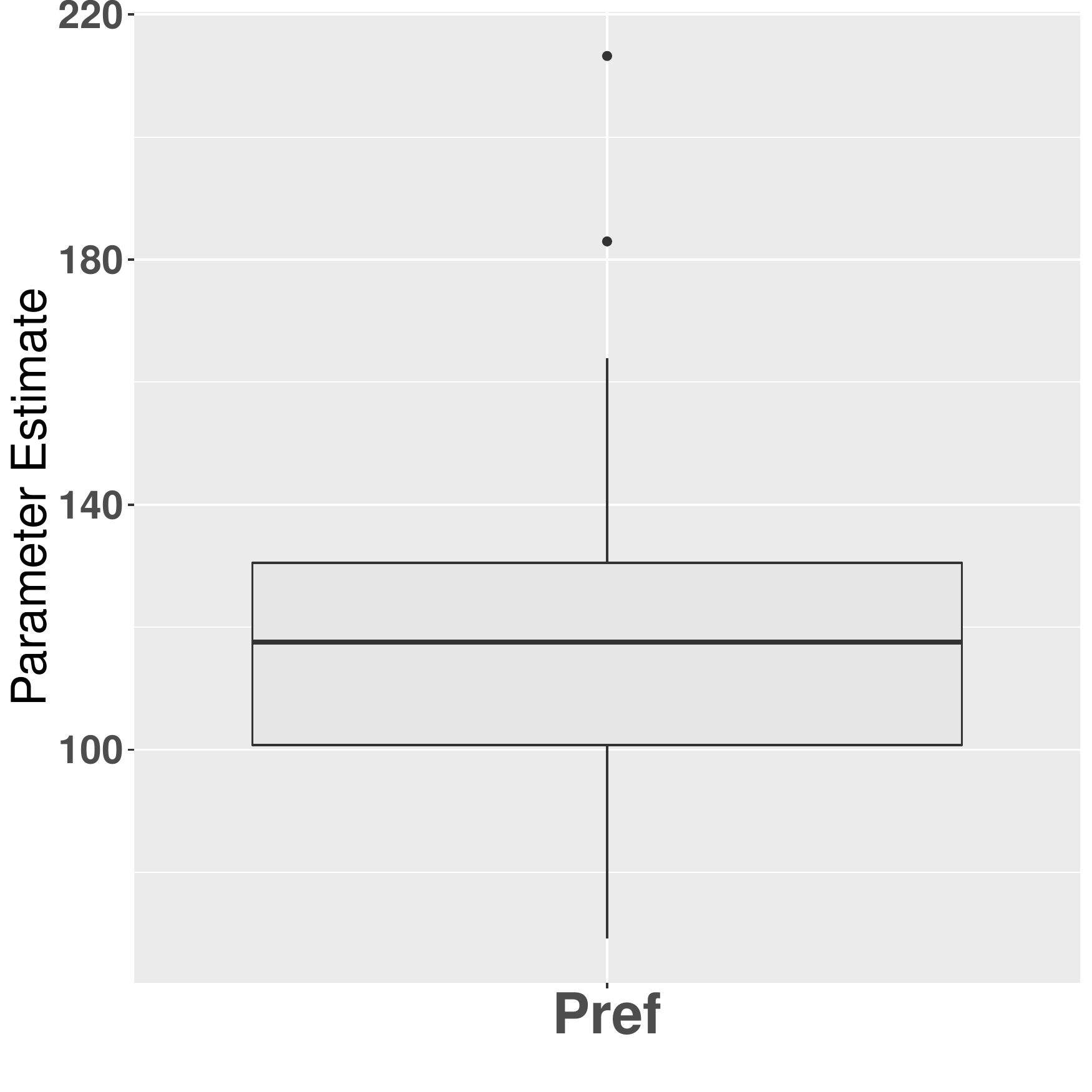}}
\subfigure[$\operatorname{Diag}(\boldsymbol{\beta}_x)$]{\includegraphics[width=2.3in]{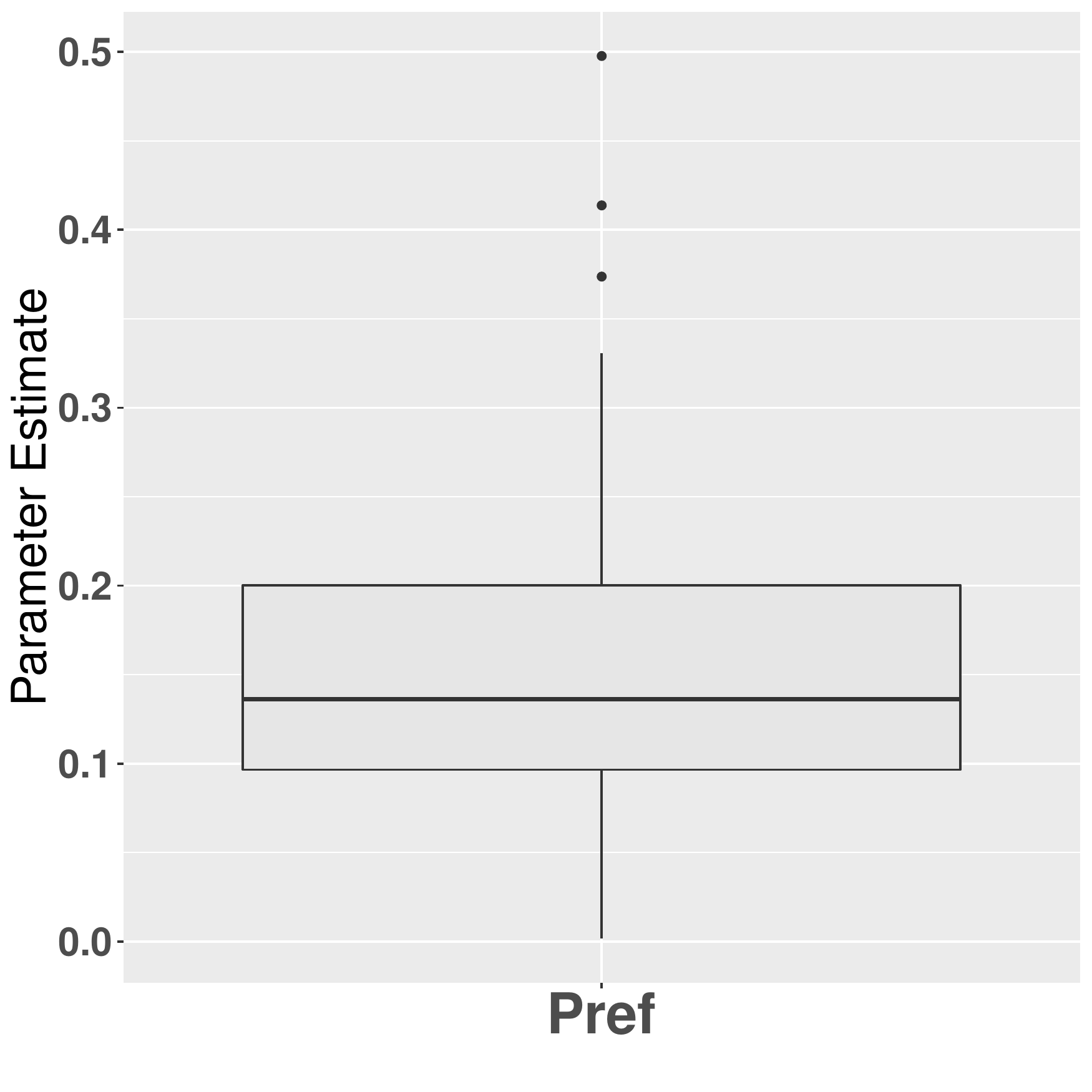}}
\subfigure[Diffusion Variance ($\operatorname{Diag}(\boldsymbol{\Sigma}_x)$)]{\includegraphics[width=2.3in]{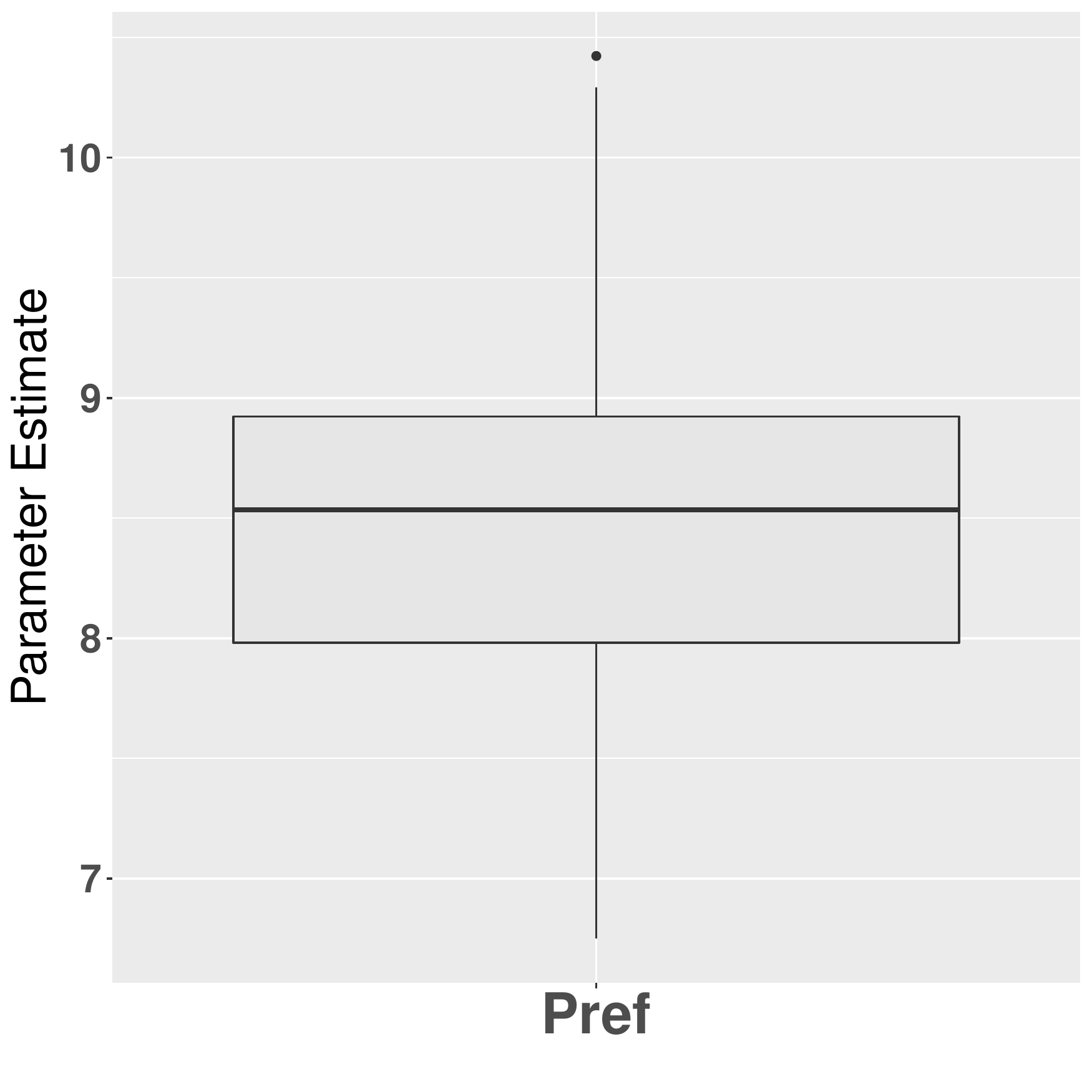}} 
\caption{Movement parameter estimates over 100 preferentially sampled simulated data sets.}
\label{fig:hybridsimmovement}
\end{figure}

For the preferentially generated data, Figure~\ref{fig:hybridsimmovement} shows that 
the movement parameters estimates, with the positive 
$\hat{\alpha}$ estimates accounting for the tendency of 
the sampler to avoid warmer warmers, which explains the
correction to the mean parameter estimates observed in Figure~\ref{fig:hybridsimfield}. 
Although these estimates cannot in general be compared with the values used to 
generate the fine-time-scale data~\citep{Gurarie2017}, the boxplots in 
 Figure~\ref{fig:hybridsimmovement} show that the estimators have a reasonable 
 sampling distribution which seem to be unimodal and symmetric around their means. 
 

\subsection{Results for Non-Preferential Data}\label{appendix:nonpref}

\begin{figure}
\centering     
\subfigure[Mean ($\mu$)]{\includegraphics[width=2.3in]{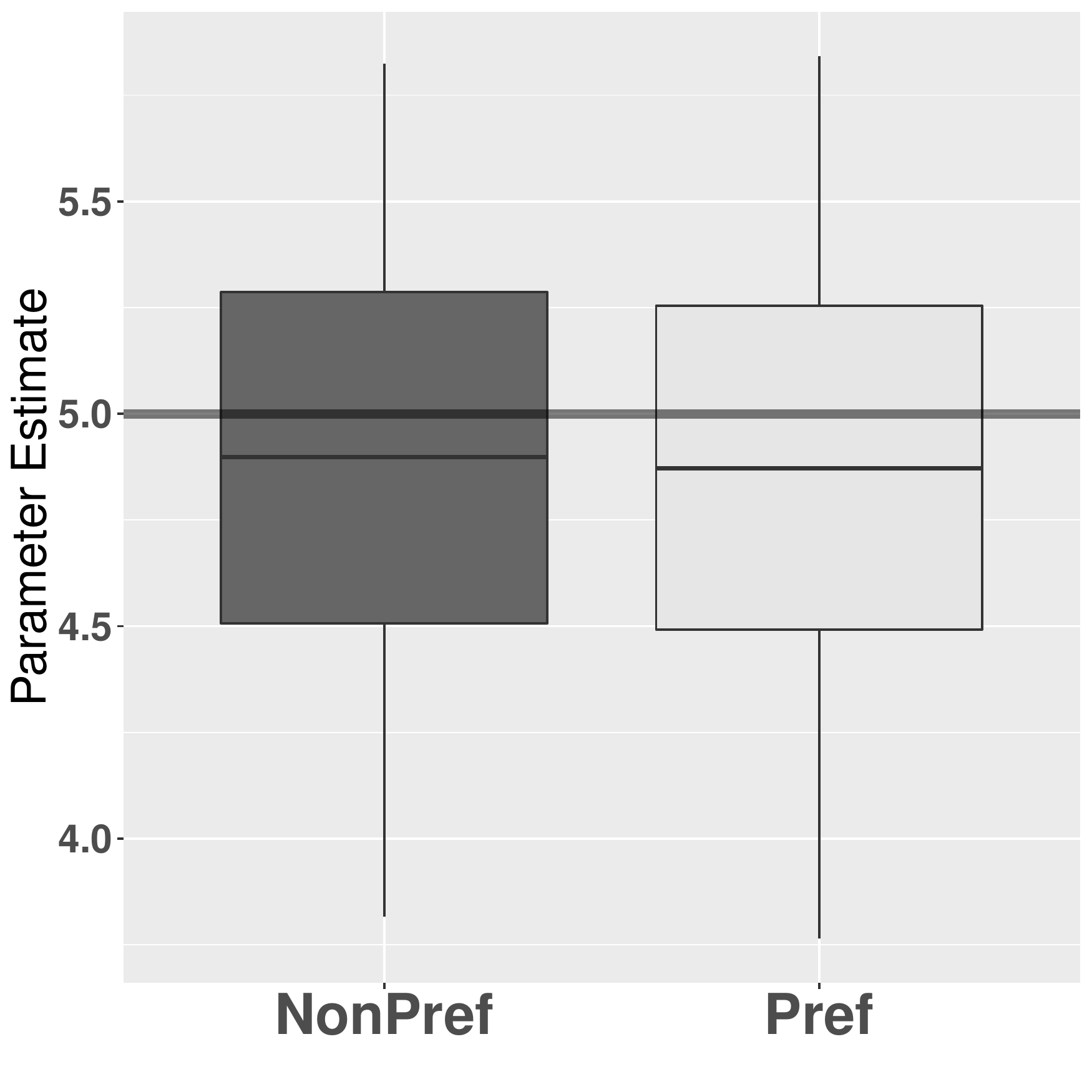}}
\subfigure[Scale ($\phi$)]{\includegraphics[width=2.3in]{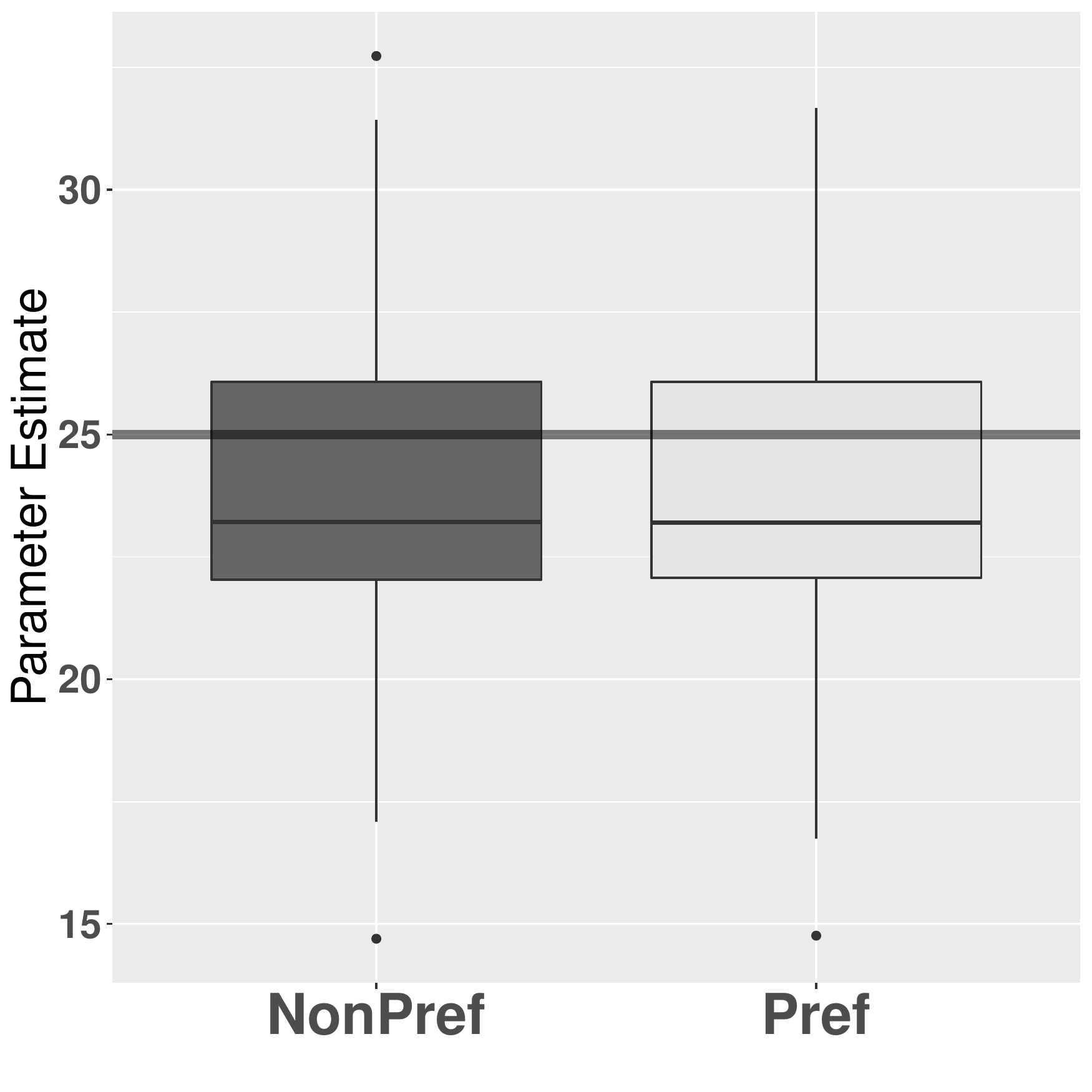}}
\subfigure[Variance ($\sigma^2$)]{\includegraphics[width=2.3in]{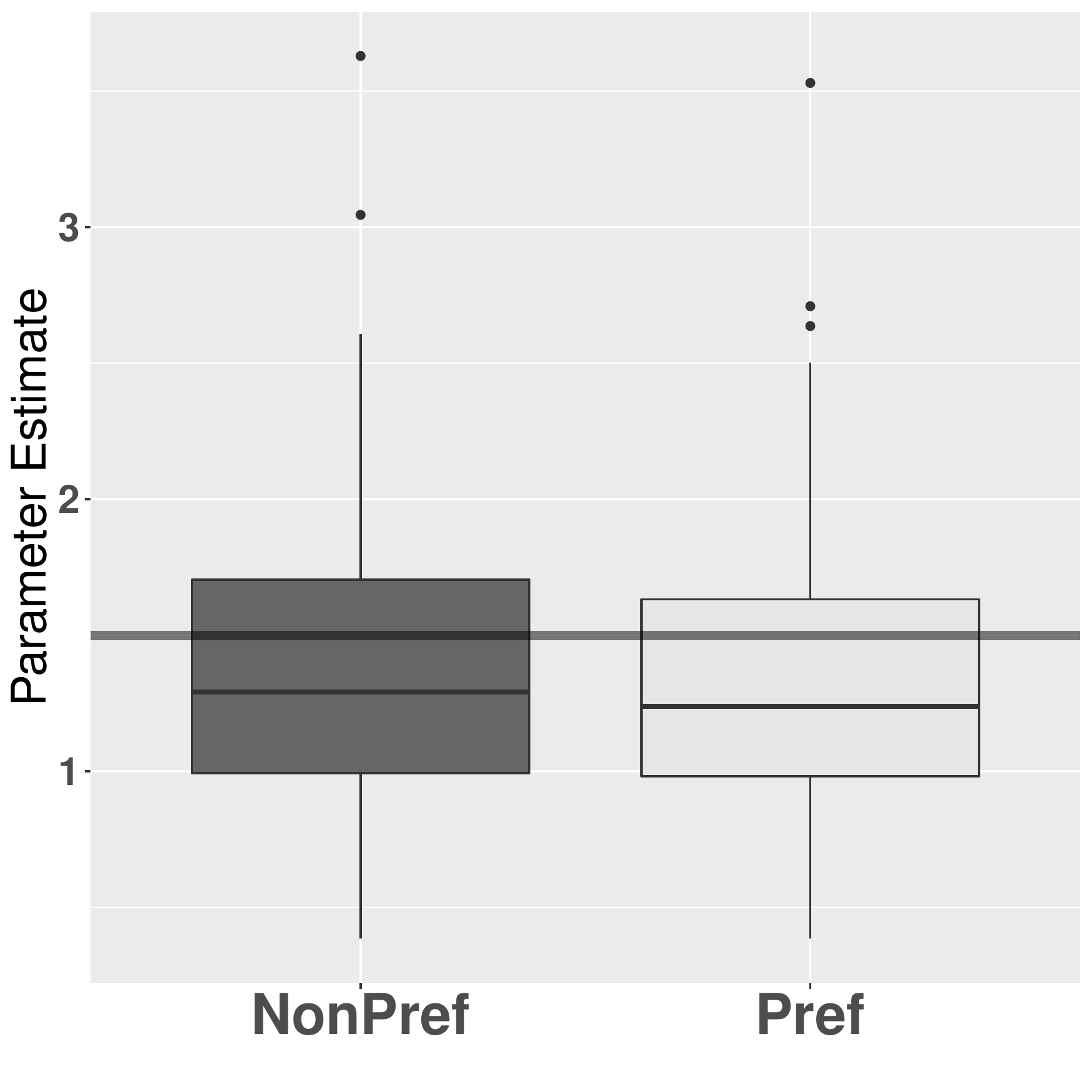}} 
\caption{Field parameter estimates over 100 non-preferentially sampled simulated data sets with true parameter values marked as grey lines. The abbreviations \texttt{NonPref} and \texttt{Pref} stand for the standard MLE (non-preferential) estimation and the one using the 
Preferential-CRW model of Section \ref{sec:CTCRWmarinemammals}.}
\label{fig:hybridsimfieldNP}
\end{figure}

We also generated 100 data sets with non-preferentially sampled data
(i.e. setting $\alpha = 0$ in \eqref{eq:expecteddrift}), 
and 
estimated the parameters of the spatial 
process using the Preferential-CRW likelihood and the 
standard one (that conditions on the observed locations). The
results in Figure \ref{fig:hybridsimfieldNP} show that, as
expected, there is no practical difference between the parameter estimates 
obtained using either likelihood when no preferential sampling is present. 

\begin{figure}
\centering     
\subfigure[$\operatorname{RMSPE}^{\text{Diff}}_i$]{\includegraphics[width=2.5in]{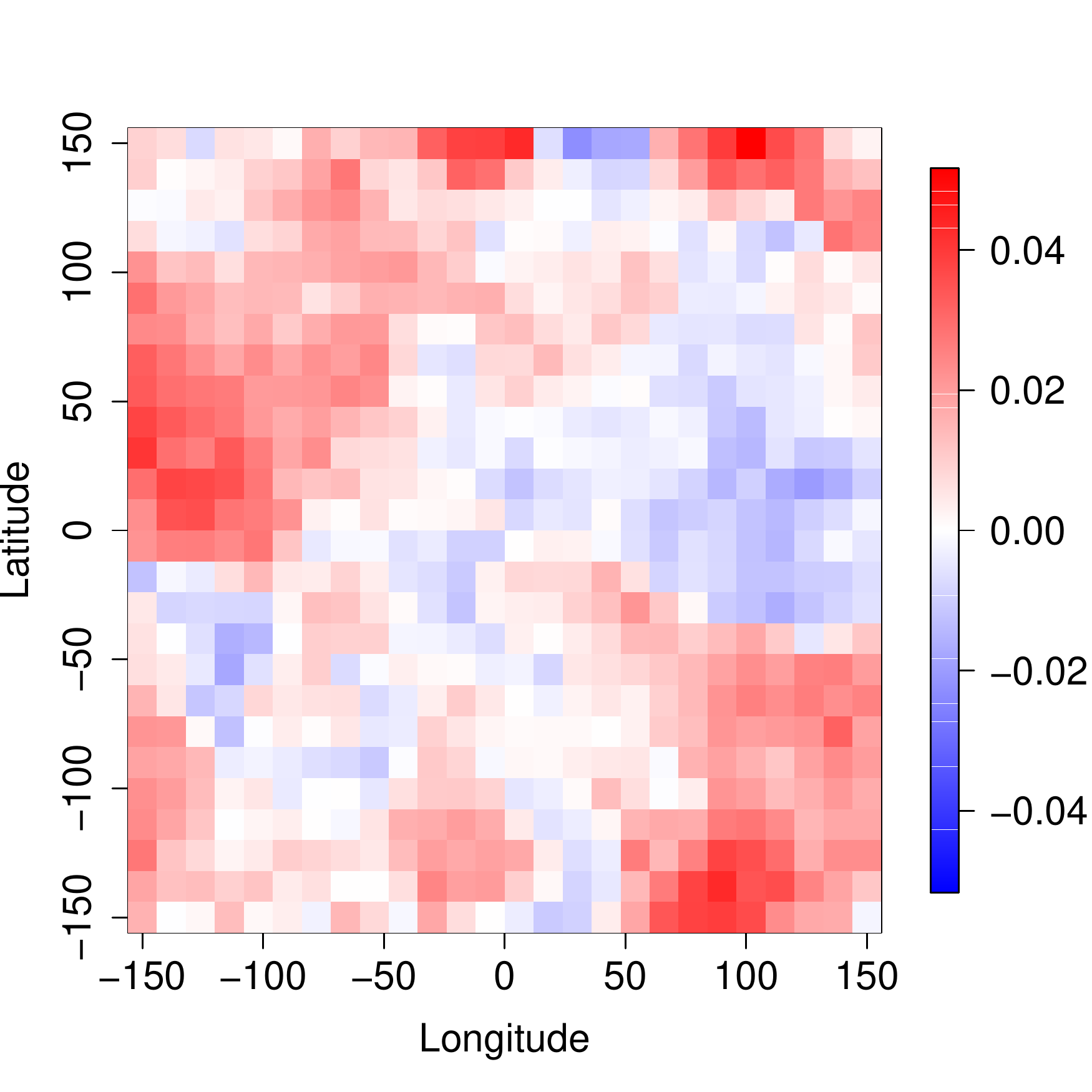}}
\subfigure[$\operatorname{MIGN}^{\text{Diff}}_j$]{\includegraphics[width=2.2in]{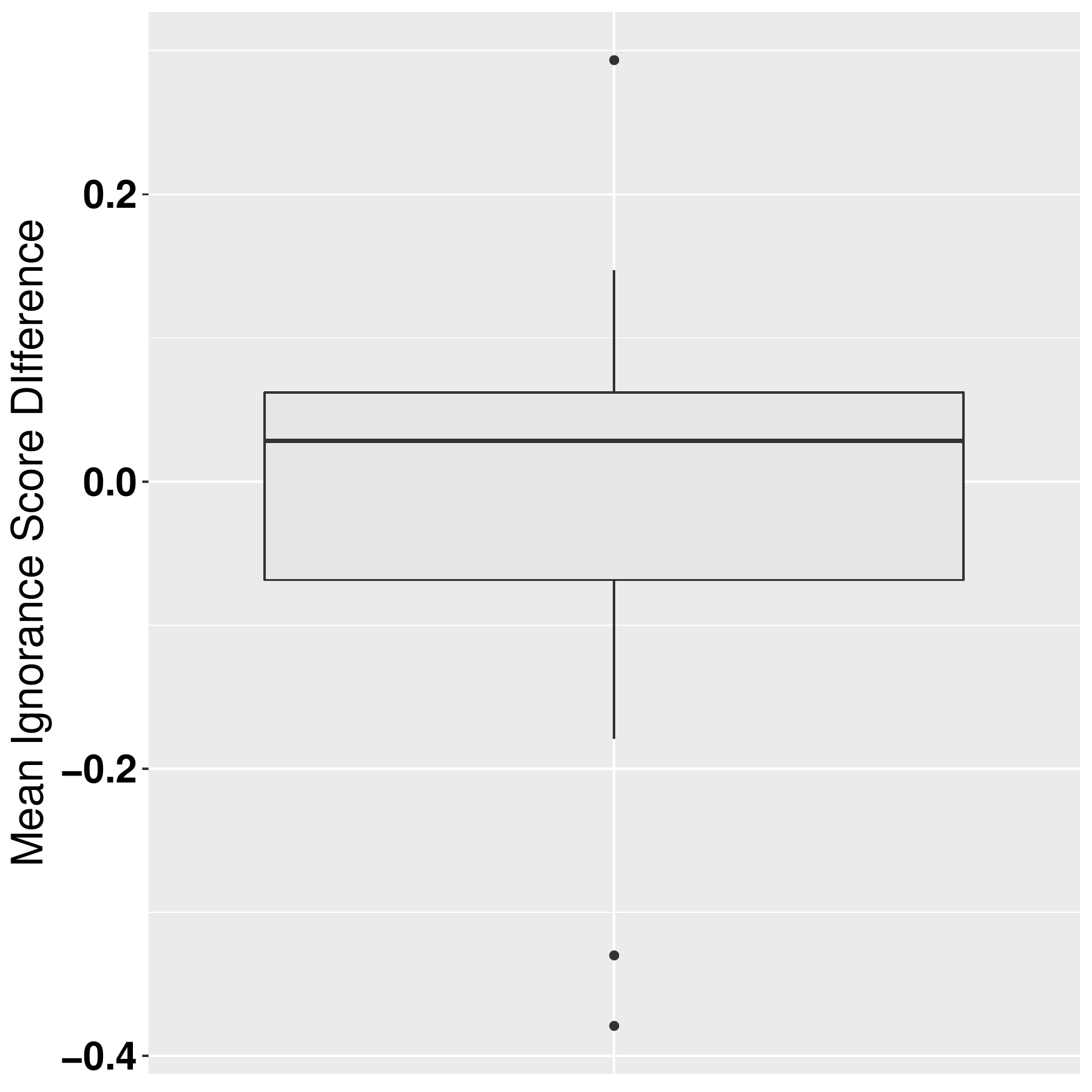}}
\subfigure[$\operatorname{LIGN}^{\text{Diff}}_i$]{\includegraphics[width=2.5in]{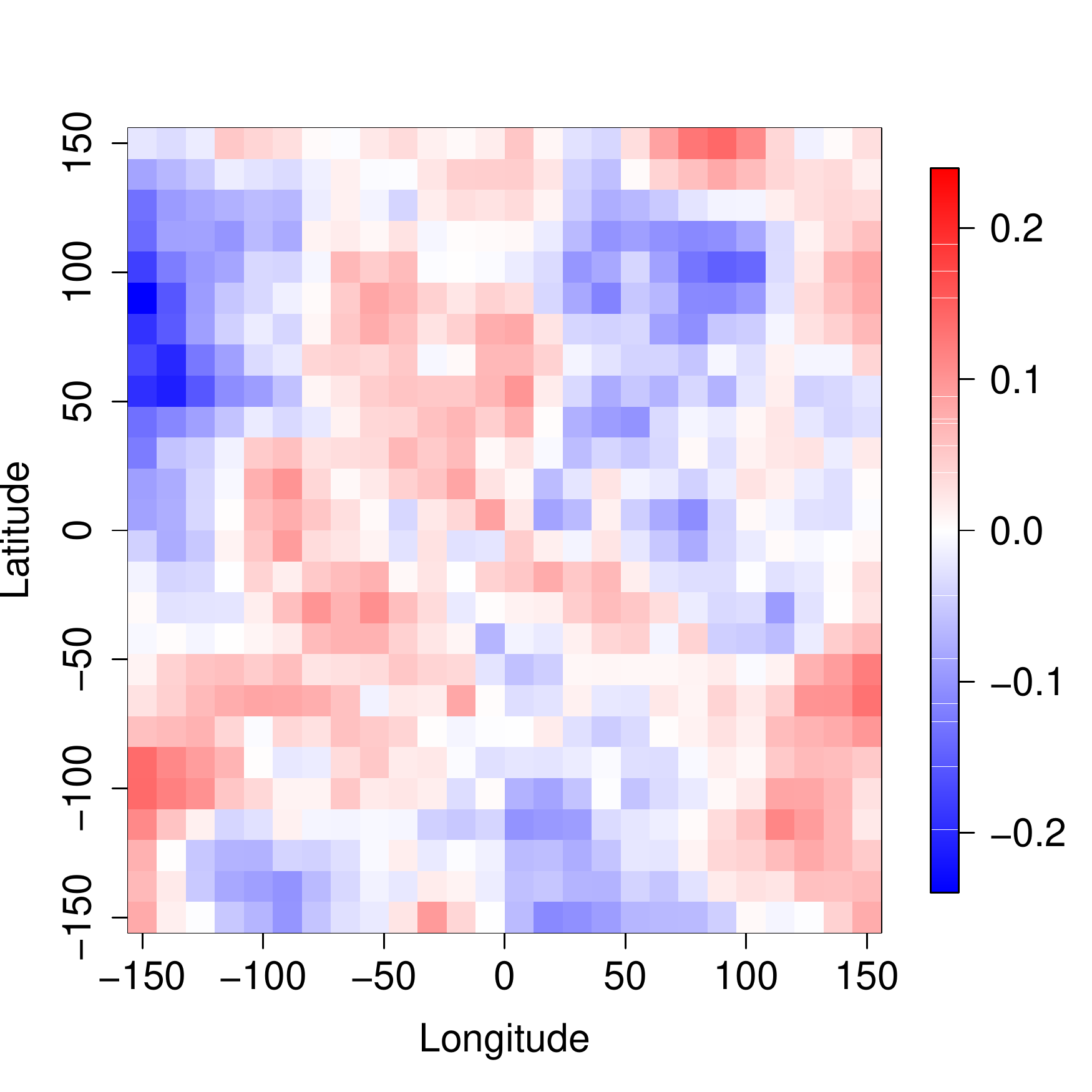}}
\caption{Comparison of Root Mean Squared Prediction Error (RMSPE) difference, Mean Ignorance Score (MIGN) difference and Location Ignorance Score (LIGN) difference respectively across 100 
non-preferentially sampled simulated data sets.}
\label{fig:hybridPredictionNP}
\end{figure}

We also performed the same comparisons as above over 
the 100 non-preferentially sampled data sets to verify that in this case, 
as expected, there was little qualitative difference between the predictions obtained using either method.
The results are displayed in Figure~\ref{fig:hybridPredictionNP} and show that, although RMSPE's are often larger with the preferential model, the difference between the two models is minor compared to the preferentially sampled data. This can be observed by comparing the scale of the differences in the plots, which are considerably smaller in the non-preferentially sampled analysis.

\section{Magnitude of Prediction Differences }\label{appendix:predif}

\begin{figure}
\centering     
\subfigure[1st Quartile of prediction differences]{\includegraphics[width=2.3in]{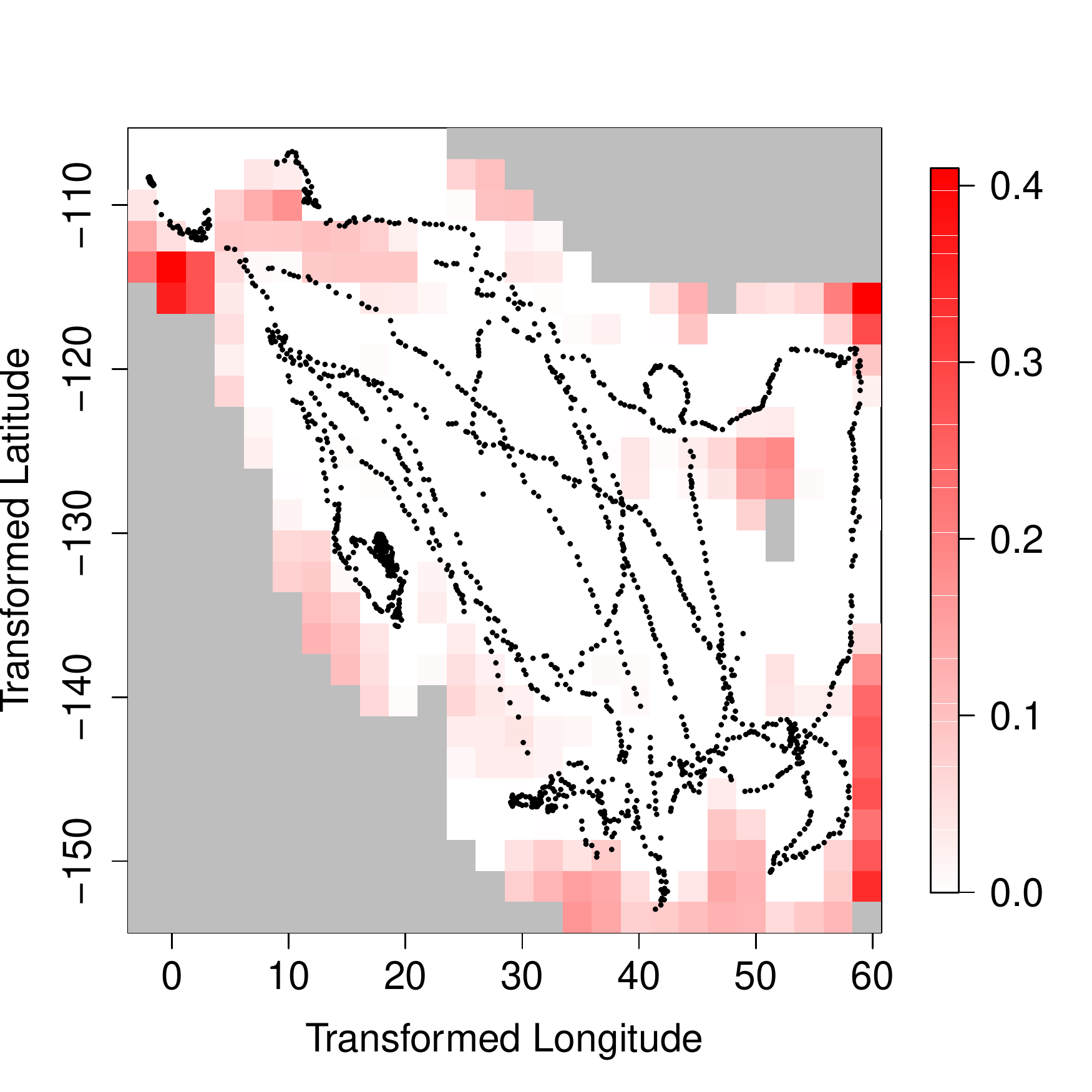}}
\subfigure[3rd Quartile of prediction differences]{\includegraphics[width=2.3in]{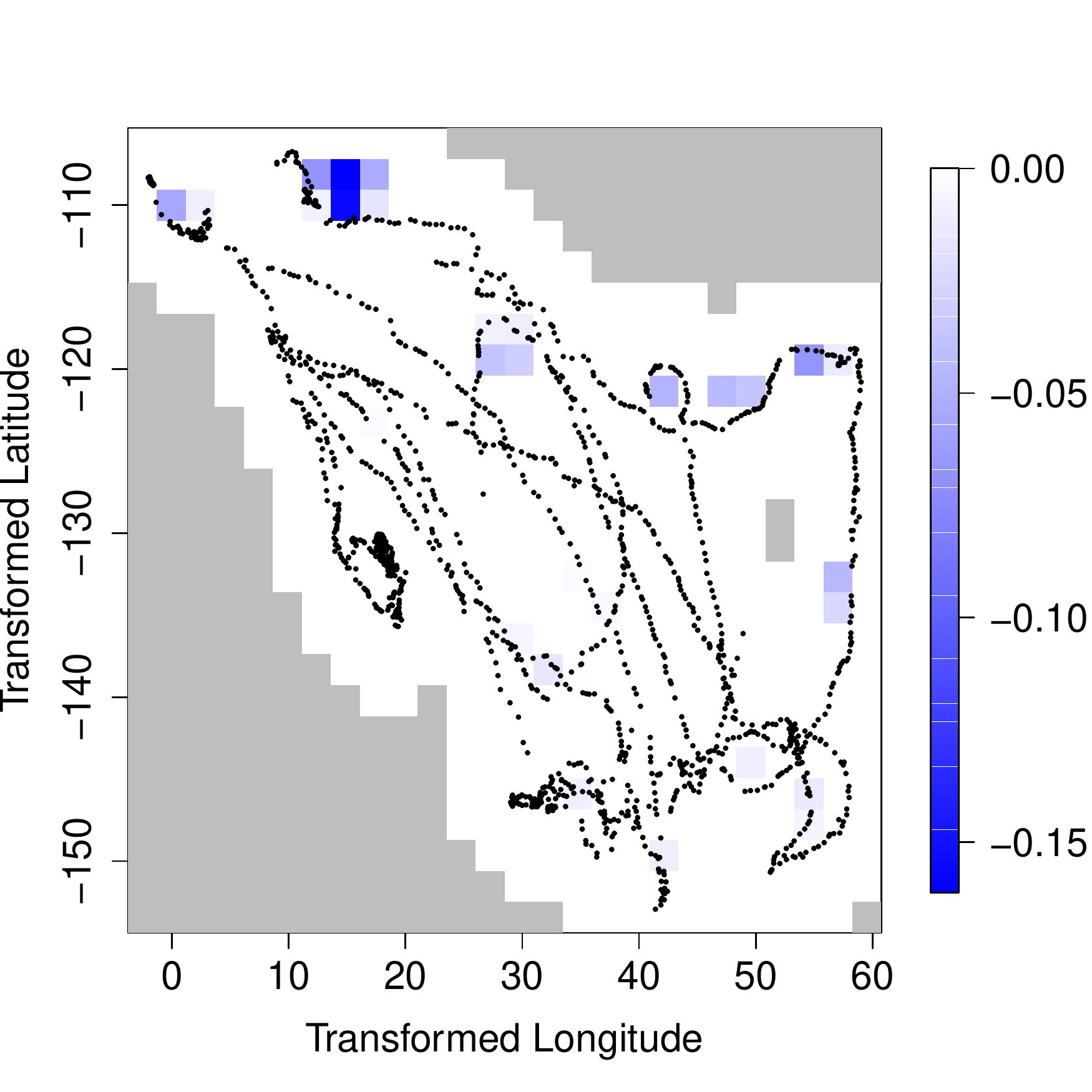}}
\caption{Quartiles of prediction differences over the 50 subsamples for the real data set analysis.}
\label{fig:qpredif}
\end{figure}

We have tried different ways to assess the 
significance of the prediction differences
in Figure~\ref{fig:RoquetPredComp}. Panel (b) shows the median difference, while the
first (25\%) and 3rd (75\%) quartiles of
the prediction differences (over the 50 subsamples) are shown in panels (a) and (b) of Figure \ref{fig:qpredif}. Note that in this plot, the white areas are locations of no-significance (negative valued regions in the 25\% quantile and positive regions in the 75\% quantile). In other words, the coloured regions of the 25\% quantile plot are regions in which prediction difference were consistently positive, and coloured regions in the 75\% quantile plot consistently negative.

As it may be expected, in large parts of the domain the differences in predictions do not show a 
clear trend either way (white locations), but some conclusions could be drawn from these figures. The coloured regions in panel (a) 
are zones were most of the predictions for the preferential sampling model were higher than from the
standard one. We see that they are in regions outside the area covered by the data, which is 
consistent with the positive estimates for $\alpha$ (see Figure 8 on page 24). Similarly, the coloured regions
in panel (b) are zones where the predictions were mostly lower. Note that these areas are almost all 
among the observed locations, which is also consistent with our conclusions above. 

\begin{supplement}[id=suppA]
  \sname{Supplement A}
  \stitle{Simulation Code}
  \slink[url]{https://github.com/msalibian/PreferentialMovement.git}
  \sdescription{Code used for the simulations in this paper, with an example shown in the \texttt{README}.}
\end{supplement}
\newpage
\bibliographystyle{imsart-nameyear}
\bibliography{AOAS1217Biblio}

\end{document}